\documentstyle[aps,eqsecnum,harvard,floats]{revtex}

\newcommand{\be}{\begin{equation}}
\newcommand{\ee}{\end{equation}}
\def\ba{\begin{eqnarray}}
\def\ea{\end{eqnarray}}
\def\go{\mathrel{\raise.3ex\hbox{$>$}\mkern-14mu\lower0.6ex\hbox{$\sim$}}}
\def\lo{\mathrel{\raise.3ex\hbox{$<$}\mkern-14mu\lower0.6ex\hbox{$\sim$}}}
\def\eps {{\varepsilon}}
\def\bB {{\bf B}}
\def\rH {{\rm H}}

\def\rkeV {{\rm keV}}
\def\rcm {{\rm cm}}

\def\br{{\bf r}}

\def\bv{{\bf v}}
\def\bR {{\bf R}}

\def\bK {{\bf K}}
\def\bp {{\bf p}}

\def\bPi {{\bf \Pi}}
\def\cE {{\cal E}}

\def\cE {{\cal E}}

\def\bsigma{{\mbox{\boldmath $\sigma$}}}
\def\etal {{\it et al.}}

\citationstyle{dcu}

\begin{document}

\title{Matter in Strong Magnetic Fields}
\author{Dong Lai}
\address{Center for Radiophysics and Space Research, Department of Astronomy,
Space Sciences Building\\
Cornell University, Ithaca, NY 14853~\footnote{Electronic address:
dong@astro.cornell.edu}}
\maketitle

\begin{abstract}
The properties of matter are significantly modified by 
strong magnetic fields, $B\gg m_e^2e^3c/\hbar^3=2.35\times 10^9\,$Gauss
($1\,{\rm G}=10^{-4}\,{\rm Tesla}$), 
as are typically found on the surfaces of
neutron stars. In such strong magnetic fields,
the Coulomb force on an electron acts as a small perturbation compared to the
magnetic force. The strong field condition can also be mimicked in 
laboratory semiconductors. Because of the strong magnetic 
confinement of electrons perpendicular to the field, atoms attain a 
much greater binding energy compared to the zero-field case, 
and various other bound states become possible, including 
molecular chains and three-dimensional condensed matter. This article
reviews the electronic structure of atoms, molecules and bulk matter,
as well as the thermodynamic properties of dense plasma, in strong
magnetic fields, $10^9\,{\rm G}\ll B\lo 10^{16}\,$G. 
The focus is on the basic physical
pictures and approximate scaling relations, although various theoretical
approaches and numerical results are also discussed. 
For the neutron star surface composed of light elements such as 
hydrogen or helium, the outermost layer constitutes a nondegenerate, partially
ionized Coulomb plasma if $B\ll 10^{14}\,$G, and may be in the form of a
condensed liquid if the magnetic field is stronger (and temperature $\lo
10^6$~K). For the iron surface, the outermost layer of the neutron star can be
in a gaseous or a condensed phase depending on the cohesive property of
the iron condensate.
\end{abstract}
\tableofcontents

\section{Introduction}
\label{I}

An electron in a uniform magnetic field $B$ gyrates in a circular orbit with
radius $\rho=m_ecv/(eB)$ at the cyclotron frequency $\omega_{ce}=eB/(m_ec)$,
where $v$ is the velocity perpendicular to the magnetic field. 
In quantum mechanics, this transverse motion is quantized into Landau levels.
The cyclotron energy (the Landau level spacing) of the electron is 
\be
\hbar\omega_{ce}=\hbar{eB\over m_e c}=11.577\,B_{12}~\rkeV,
\ee
and the cyclotron radius (the characteristic size of the wave packet) becomes
\be
\hat\rho= \left({\hbar c\over eB}\right)^{1/2}=2.5656\times
10^{-10}B_{12}^{-1/2}~\rcm,
\ee
where $B_{12}=B/(10^{12}\,{\rm G})$ 
is the magnetic field strength in units of $10^{12}$~Gauss,
a typical field found on the surfaces of neutron stars (see Sec.~I.A). 
When studying matter in magnetic fields, the natural (atomic) unit
for the field strength, $B_0$, is set by 
$\hbar\omega_{ce}=e^2/a_0$, or equivalently by $\hat\rho=a_0$, where
$a_0$ is the Bohr radius. Thus it is convenient to 
define a dimensionless magnetic field strength $b$ via 
\be
b\equiv {B\over B_0};\qquad B_0={m_e^2e^3c\over\hbar^3}=2.3505\times 10^9\,
{\rm G}.
\label{eqb0}\ee
For $b\gg 1$, the electron cyclotron energy 
$\hbar\omega_{ce}$ is much larger than the typical Coulomb energy,
so that the properties of atoms, molecules and condensed matter are 
qualitatively changed by the magnetic field.\footnote{Note that this statement 
applies to individual atoms, molecules and zero-pressure condensed matter.
In a medium, when the density or temperature is sufficiently
high, the magnetic effects can be smeared out even for $b\gg 1$; see 
Sec.~VI. The strong field condition is also modified by the ion charge;
see Sec.~III.C.}
In such a strong field regime, the usual perturbative
treatment of the magnetic effects (e.g., Zeeman splitting of 
atomic energy levels) does not apply
(see Garstang 1977 for a review of atomic physics at $b\lo 1$).
Instead, the Coulomb forces act as a perturbation to the magnetic
forces, and the electrons in an atom settle into the ground Landau level. 
Because of the extreme confinement ($\hat\rho\ll a_0$) of the electrons in the
transverse direction (perpendicular to the field), the Coulomb force becomes
much more effective in binding the electrons along the magnetic field
direction. The atom attains a cylindrical structure. Moreover, it is possible
for these elongated atoms to form molecular chains by covalent bonding along
the field direction. Interactions between the linear chains can then lead to
the formation of three-dimensional condensates. The properties of atoms,
molecules and bulk matter in strong magnetic fields of $b\gg 1$ is the subject
of this review.

\subsection{Astrophysics Motivation}

Strong magnetic fields with $b\gg 1$ exist on the surfaces on 
neutron stars (NSs). Most radio pulsars and accreting NSs in
X-ray binaries have surface fields in the range of $10^{12}-10^{13}$~G;
even recycled millisecond pulsars and old NS's in low-mass X-ray
binaries have fields $B=10^8-10^9$~G (e.g., 
Lyne and Graham-Smith 1998; Lewin \etal~1995). The physical
upper limit to the NS magnetic field strength follows from the
virial theorem of magnetohydrostatic equilibrium (Chandrasekhar \& Fermi 1953;
see Shapiro and Teukolsky 1983). 
The magnetic energy of the NS (mass $M_{\rm NS}$, radius $R_{\rm NS}$),
$(4\pi R_{\rm NS}^3/3)(B^2/8\pi)$, can never exceed its gravitational binding
energy, $\sim GM_{\rm NS}^2/R_{\rm NS}$; this gives
\be
B\lo 10^{18} \left({M_{\rm NS}\over 1.4M_\odot}\right)\left({R_{\rm NS}\over
10~{\rm km}}\right)^{-2}{\rm G}.
\ee
It has been suggested that magnetic fields of
order $10^{15}$~G or stronger can be generated by dynamo processes in
proto-neutron stars (Thompson and Duncan 1993), and recent
observations (e.g., Vasisht and Gotthelf,~1997; Kouveliotou \etal,~1998,~1999;
Hurley \etal~1999; Kaspi \etal~1999; Mereghetti 2000) have lent support to
the idea that soft
gamma-ray repeaters and slowly spinning (with periods of a few seconds)
``anomalous'' x-ray pulsars in supernova remnants are NSs endowed
with superstrong magnetic fields $B\go 10^{14}$~G, the so-called ``magnetars'' 
(Duncan and Thompson 1992; Paczy\'nski 1992;
Thompson and Duncan,~1995,~1996). Finally, magnetism has been detected in 
a few dozen white dwarfs (out of a total of about 2000) in the range from
$10^5$~G to $10^9$~G (e.g., Koester and Chanmugan 1990; Jordan 1998;
Wickramasinghe and Ferrario 2000).

The main astrophysical motivation to study matter in strong magnetic fields
arises from the importance of understanding NS surface layers,
which play a key role in many NS processes and observed phenomena.
Theoretical models of pulsar and magnetar magnetospheres depend on 
the cohesive properties of the surface matter in strong
magnetic fields (e.g., Ruderman and Sutherland 1975; Michel 1991;
Usov and Melrose 1996; 
Zhang and Harding 2000). More importantly, the surface layer directly mediates
the thermal radiation from the NS. It has long been recognized that NSs are
sources of soft X-rays during the $\sim 10^5-10^6$ years of cooling phase after
their birth in supernova explosions (Chiu and Salpeter 1964; Tsuruta 1964;
Bahcall and Wolf 1965). The cooling
history of the NS depends on poorly-constrained interior physics,
such as nuclear equation of state, superfluidity and internal magnetic 
fields (see, e.g., Pethick 1992, Page 1998, Tsuruta 1998, 
Prakash \etal~2000; Yakovlev \etal~2001 for review). 
The advent of imaging X-ray telescopes in recent years has now
made it possible to observe isolated NSs directly by their surface
radiation. In particular, recent X-ray observatories such as ROSAT 
have detected pulsed X-ray thermal emission from a number of radio pulsars 
(see Becker and Tr\"umper 1997, Becker 2000, Becker \&
Pavlov 2001 for review). 
Some of the observed X-rays are likely to be
produced by nonthermal magnetospheric emission, but at least three pulsars (PSR
B1055-52, B0656+14, Geminga) show emission dominated by a thermal component
emitted from the whole NS surface, with temperatures in the range
$(2-10)\times 10^5$~K. A few of the X-ray emitting radio pulsars show
thermal-like radiation of higher temperatures, in the range $(1-5)\times
10^6$~K, from an area much smaller than that of the stellar surface, 
indicating a hot polar cap on the NS surface. Several nearby pulsars have also
been detected in the extreme ultraviolet (Edelstein, Foster and Bowyer 1995; 
Korpela and Bowyer 1998)
and in the optical band (e.g., Pavlov \etal~1997; Caraveo \etal~2000)
with spectra consistent with thermal radiation from NS surfaces.
On the other hand, old isolated NSs ($10^8-10^9$ of 
which are thought to exist in the Galaxy), heated through accretion 
from interstellar material, are also expected to be common sources
of soft X-ray/EUV emission (e.g, Treves and Colpi 1991; 
Blaes and Madau 1993; Treves \etal~2000). 
Several radio-quiet isolated accreting NSs have been
detected in the X-ray and optical band (e.g.,
Walter, Wolk and Neuh\"auser 1996; Caraveo, Bignami and Tr\"umper 1996; 
Walter and Matthews 1997).
Finally, the quiescent X-ray emissions from soft gamma-ray repeaters 
and anomalous X-ray pulsars may be powered by the internal heating 
associated with decaying magnetic fields (Thompson and Duncan 1996).
The X-rays originate near the stellar surface, and therefore allow one 
to probe radiative transport in the superstrong field regime. 
Observations indicate that some of the sources (particularly anomalous 
X-ray pulsars) have a thermal component in their X-ray spectra (e.g.,
Mereghetti 2000).
Overall, the detections of surface emission from NSs
can provide invaluable information on the structure and evolution 
of NSs, and enable one to put constraints on the nuclear 
equation of state, various heating/accretion processes, magnetic field
structure and surface chemical composition. 
The recently launched X-ray telescopes, including Chandra X-ray 
Observatory and XMM-Newton Observatory, have much improved 
sensitivity and spectral resolution in the soft X-ray band, 
making it promising for spectroscopic studies of isolated or 
slowly-accreting NSs. Since the surface layer/atmosphere directly determines
the characteristics of the thermal emission, proper interpretations of
the observations require a detailed understanding of the physical properties of
the NS envelope in the presence of intense magnetic fields ($B\go 10^{12}$~G).

\subsection{Laboratory Physics Motivation}

The highest static magnetic field currently produced in a terrestrial
laboratory is $45$~Tesla ($4.5\times 10^5$~G), far below $B_0$;
stronger transient field of order $10^3$~Tesla can be produced using 
explosive flux compression techniques, but this is still below $B_0$
(e.g., Crow \etal~1998). However, high magnetic-field conditions can be
mimicked in some semiconductors where a small effective electron mass
$m_{\ast}$ and a large dielectric constant $\varepsilon$ 
reduce the Coulomb force relative to the magnetic force. For
hydrogen-like excitons in semiconductors, the atomic unit of length
is $a_{0\ast}=\varepsilon\hbar^2/(m_\ast e^2)$, and the corresponding
natural unit for magnetic field is $B_{0\ast}={m_{\ast}^2e^3c/
({\varepsilon}^2\hbar^3})$. For example, in GaAs, the dielectric constant is
$\varepsilon=12.56$, and the electron bound to a positively charged donor has
an effective mass $m_{\ast}=0.00665m_e$, thus $B_{0\star}=6.57$~Tesla. The
critical field $B_{0\ast}$ can be as small as $0.2$~Tesla for InSb.
Such a low value of $B_{0\ast}$ implies that the structure of excitons, 
biexcitonic molecules and quantum dots (which resemble multi-electron
atoms; see Kastner 1992) in semiconductors must experience significant changes
already in laboratory magnetic fields (e.g., Timofeev and Chernenko 1995; 
Lieb \etal~1995; Klaassen \etal~1998). Some of the 
earliest studies of atoms in superstrong magnetic fields (Elliot 
and Loudon 1960; Hasegawa and Howard 1961) were motived by applications
in semiconductor physics.

\subsection{Plan of This Paper}

In this paper we review the properties of different forms of
matter (atoms, molecules and bulk condensed matter) in strong magnetic fields. 
We also discuss astrophysical situations where magnetized matter 
plays an important role. 
We shall focus on magnetic field strengths in the range of $B\gg
10^{9}$~G so that $b\gg 1$ is well-satisfied, although in several places
(Sec.~III.D and Sec.~IV.E) we will also touch upon issues for $b\go 1$. 
Throughout the paper, our emphasis is on physical understanding and 
analytic approximate relations (whenever they exist), 
rather than on computational techniques for the electronic
structure calculations, although we will discuss the later aspects and provide
pointers to the literature. 

This paper is organized as follows. After a brief summary of
the basics of electron Landau levels in Section II, we discuss the 
physics of various bound states in strong magnetic fields
in Section III (Atoms), Section IV (Molecules) and Section V 
(Condensed Matter). Section VI summarizes the thermodynamic
properties of free electron gas (at finite density and temperature). 
In Section VII we review the physical properties of the envelope of
a strongly magnetized neutron star. 

Throughout the paper we shall use real physical units and 
atomic units (a.u.) interchangeably, whichever is more convenient. 
Recall that in atomic units, mass and length are expressed in units 
of the electron mass $m_e$ and the Bohr radius 
$a_0=0.529\times 10^{-8}$ cm, energy in units of 
$2~{\rm Ryd}=e^2/a_0=2\times 13.6$~eV; field strength 
in units of $B_0$ [Eq.~(\ref{eqb0})], temperature in units of 
$3.16\times 10^5$~K, and pressure in units of $e^2/a_0^4=
2.94\times 10^{14}$~dynes~cm$^{-2}$.

\subsection{Bibliographic Notes}

Theoretical research on matter in superstrong magnetic fields started
in the early sixties. A large number of papers have been written 
on the subject over the years and they are scattered in astrophysics
atomic/molecular physics and condensed matter physics literatures. 
Although we have tried to identify original key papers whenever possible, 
our references put more emphasis on recent works from which earlier 
papers can be found. We apologize to the authors of the relevant papers
which are not mentioned here.

In recent years there has been no general review article
that covers the broad subject of matter in strong magnetic fields,
although good review articles exist on aspects of the problem.
Extensive review of atoms (especially for H and He)
in strong magnetic fields, including tabulations of numerical results,
can be found in the monograph ``Atoms in Strong Magnetic Fields'' by Ruder et
al.~(1994). A recent reference is the conference proceedings on 
``Atoms and Molecules in Strong Magnetic Fields'' edited by 
Schmelcher and Schweizer (1998). An insightful, short review of earlier 
works is Ruderman (1974). Other reviews on general physics in strong magnetic
fields include Canuto and Ventura (1977) and the monograph by Meszaros (1992).

\section{Basics on Landau Levels}
\label{CBI}

The quantum mechanics of a charged particle in a magnetic field
is presented in many texts (e.g., Landau and Lifshitz 1977; 
Canuto and Ventura 1977; Sokolov and Ternov 1968; M\'esz\'aros 1992). Here we
summarize the basics needed for our later discussion. 

For a free particle of charge $e_i$ and mass $m_i$ in a constant
magnetic field (assumed to be along the $z$-axis), the kinetic
energy of transverse motion is quantized into Landau levels
\be
E_\perp={1\over 2}m_i\bv_\perp^2={1\over 2m_i}\bPi_\perp^2\rightarrow 
\left(n_L+{1\over 2}\right)\hbar\omega_c,\qquad n_L=0,1,2,\cdots
\ee
where $\omega_c=|e_i|B/(m_ic)$ is the cyclotron (angular) frequency, 
$\bPi={\bf P}-(e_i/c){\bf A}=m_i{\bf v}$ is the
mechanical momentum, ${\bf P}=-i\hbar{\bf\nabla}$ is the 
canonical momentum, and ${\bf A}$ is the vector potential of the magnetic
field. 

A Landau level is degenerate, reflecting the fact that 
the energy is independent of the location of the guiding center of the
gyration. To count the degeneracy, it is 
useful to define the {\it pseudomentum} (or the generalized momentum)
\be
{\bf K}=\bPi+(e_i/c){\bf B}\times {\bf r}.
\ee
That $\bK$ is a constant of motion (i.e., it commutes with the Hamiltonian)
can be easily seen from the classical equation of motion for the particle,
$d\bPi/dt=(e_i/c)(d\br/dt)\times\bB$. 
Mathematically, the conservation of $\bK$ 
is the result of the invariance of the Hamiltonian under a
spatial translation plus a gauge transformation (Avron \etal~1978).
The parallel component $K_z$
is simply the linear momentum, while the constancy of the perpendicular
component $\bK_{\perp}$ is the result of the fact that the guiding center of
the gyro-motion does not change with time. The position vector
${\bf R}_c$ of this guiding center is related to $\bK_{\perp}$ by
\be
\bR_c={c\bK_{\perp}\times {\bf B}\over e_iB^2}={c\over e_iB}\bPi_\perp
\times\hat\bB+\br_\perp,
\label{eqbrc}\ee
($\hat\bB$ is the unit vector along $\bB$).
Clearly, the radius of gyration, $\rho=m_icv_\perp/(|e_i|B)$, is quantized
according to
\be
|\br_\perp-\bR_c|={c\over |e_i|B}|\bPi_\perp|
\rightarrow (2n_L+1)^{1/2}\hat\rho,
\label{eqrhon}\ee
where 
\be
\hat\rho=\left({\hbar c\over |e_i|B}\right)^{1/2}=b^{-1/2}
\ee
is the cyclotron radius (or the magnetic length). 
We can use $\bK$ to classify the eigenstates. However,
since the two components of $\bK_{\perp}$ do not commute,
$[K_x,K_y]=-i\hbar (e_i/c) B$, only one of the components can be diagonalized
for stationary states. This means that the guiding center
of the particle can not be specified.
If we use $K_x$ to classify the states,
then the wavefunction has the well-known form $e^{K_x x/\hbar}\phi(y)$
(Landau and Lifshitz 1977), where 
the function $\phi(y)$ is centered at $y_c=-cK_x/(e_iB)$ [see
Eq.~(\ref{eqbrc})]. The Landau degeneracy in 
an area ${\cal A}_g=L_g^2$ is thus given by 
\be
{L_g\over h}\int\! dK_x={L_g\over h}|K_{x,g}|={\cal A}_g
{|e_i|B\over h c}={{\cal A}_g\over 2\pi\hat\rho^2},
\label{eqdeg}\ee
where we have used $K_{x,g}=-e_iBL_g/c$.
On the other hand, if we choose to diagonalize $K_{\perp}^2=K_x^2+K_y^2$,
we obtain the Landau wavefunction $W_{nm}(\br_\perp)$
in cylindrical coordinates (Landau and Lifshitz 1977),
where $m$ is the ``orbital'' quantum number (denoted by $s$ or $-s$
in some references). For the ground Landau level, this is
(for $e_i=-e$)
\be
W_{0m}(\br_\perp)\equiv
W_{m}(\rho,\phi)={1\over (2\pi m!)^{1/2}\hat\rho}
\left({\rho\over \sqrt{2}\hat\rho}\right)^m
\exp\left({\rho^2\over 4\hat\rho^2}\right)\exp(-im\phi),
\label{eqw0m}\ee
where the normalization $\int d^2\br_\perp\,|W_m|^2=1$ is
adopted. The (transverse) distance of the guiding center of the particle
from the origin of the coordinates is given by 
\be
|\bR_c|\rightarrow
\rho_m=(2m+1)^{1/2}\hat\rho,~~~~~m=0,1,2,\cdots
\label{eqrhom}\ee
The corresponding value of $K_{\perp}$ is 
$K_{\perp}^2=(\hbar |e_i|B/c)(2m+1)$. 
Note that $K_\perp^2$ assumes discrete values since $m$ is required to be an 
integer in order for the wavefunction to be single-valued. 
The degeneracy $m_g$ of the
Landau level in an area ${\cal A}_g=\pi R_g^2$ is then determined by
$\rho_{m_g} \simeq (2m_g)^{1/2}\hat\rho=R_g$, 
which again yields $m_g={\cal A}_g|e_i|B/(h c)$ as in Eq.~(\ref{eqdeg}).
We shall refer to different $m$-states as different Landau
orbitals. Note that despite the similarity between equations (\ref{eqrhom})
and (\ref{eqrhon}), their physical meanings are quite different:
the circle $\rho=\rho_m$ does not correspond to any 
gyro-motion of the particle, and the energy is independent of $m$.

We also note that $K_{\perp}^2$ is related to the $z$-projection of angular
momentum $J_z$, as is evident from the $e^{-im\phi}$ factor in the
cylindrical wavefunction [Eq.~(\ref{eqw0m})]. In general, we have
\ba
J_z&=&xP_y-yP_x={1\over 2e_iB}(\bK_{\perp}^2-\bPi_{\perp}^2)\nonumber\\
&=& (m-n_L) {|e_i|\over e_i},
\ea
where we have used $\bPi_{\perp}^2=(\hbar |e_i|/c)B(2n_L+1)$.

Including the spin energy of the electron ($e_i\rightarrow 
-e$, $\omega_c\rightarrow
\omega_{ce}$), $E_{\sigma_z}=e\hbar/(2m_ec)
\bsigma\cdot\bB=\hbar\omega_{ce}\sigma_z/2$, the total electron 
energy can be written as
\be
E=n_L\hbar\omega_{ce}+{p_z^2\over 2m_e},
\label{eqefree}\ee
where the index $n_L$ now includes the spin. 
For the ground Landau level ($n_L=0$),
the spin degeneracy is one ($\sigma_z=-1$); for excited levels, 
the spin degeneracy is two.  

For extremely strong magnetic fields such that $\hbar\omega_{ce}\go m_ec^2$, or
\be
B\go B_{\rm rel}={m_e^2c^3\over e\hbar}=
{B_0\over \alpha^2}= 4.414 \times 10^{13}~{\rm G},
\ee
($\alpha=e^2/\hbar c$ is the fine structure constant), 
the transverse motion of the electron becomes relativistic. Equation 
(\ref{eqefree}) for the energy of a free electron should be replaced by 
(e.g., Johnson and Lippmann 1949)
\be
E=\left[c^2p_z^2+m_e^2c^4\left(1+2n_L\beta\right)\right]^{1/2},
\label{eqrel}\ee
where
\be
\beta\equiv {B\over B_{\rm rel}}=\alpha^2\,b.
\label{eqrel2}\ee
Higher order corrections in $e^2$ to Eq.~(\ref{eqrel}) have the form
$(\alpha/4\pi)m_ec^2 F(\beta)$, with $F(\beta)=-\beta$ for $\beta\ll 1$ and
$F(\beta)=\left[\ln(2\beta)-(\gamma+3/2)\right]^2+\cdots$ for $\beta\gg 1$,
where $\gamma=0.5772$ is Euler's constant (Schwinger 1988); these corrections
will be neglected.

When studying bound states (atoms, molecules and condensed matter near zero 
pressure) in strong magnetic fields (Sec.~III-Sec.~V), we shall use 
nonrelativistic quantum mechanics, even for $B\go B_{\rm rel}$. 
The nonrelativistic treatment of bound states
is valid for two reasons:
(i) For electrons in the ground Landau level, the free electron energy
reduces to $E\simeq m_ec^2+p_z^2/(2m_e)$; the electron remains 
nonrelativistic in the $z$-direction (along the field axis) as long as the
binding energy $E_B$ is much less than $m_ec^2$; (ii) 
The shape of the Landau wavefunction in the relativistic theory 
is the same as in the nonrelativistic theory, as seen from the fact that 
$\hat\rho$ is independent of the particle mass. 
Therefore, as long as $E_B/(m_e c^2)\ll 1$, the relativistic
effect on bound states is a small correction (Angelie and Deutch 1978).
For bulk matter under pressure, the relativistic correction 
becomes increasingly important as density increases (Sec.~VI).


\section{Atoms}

\subsection{Hydrogen Atom}

In a strong magnetic field with $b\gg 1$, the
electron is confined to the ground Landau level (``adiabatic 
approximation''), and the Coulomb potential can be treated as a 
perturbation. Assuming infinite proton mass (see Sec.~III.E),
the energy spectrum of the H atom is specified by 
two quantum numbers,
$(m,\nu)$, where $m$ measures the mean transverse separation
[Eq.~(\ref{eqrhom})] between the electron and the proton,
while $\nu$ specifies the number of nodes in the $z$-wavefunction. 
We may write the electron wavefunction as 
$\Phi_{m\nu}(\br)=W_m(\br_{\perp})f_{m\nu}(z)$. 
Substituting this function into the 
Schr\"odinger equation and averaging over the transverse
direction we obtain a one-dimensional Schr\"odinger equation for $f_{m\nu}(z)$:
\be
-{\hbar^2\over 2m_e\hat\rho^2} f_{m \nu}''- {e^2 \over \hat\rho}
V_m(z)f_{m \nu}=E_{m \nu} f_{m \nu},\qquad
(m,\nu=0,1,2,\cdots)
\label{eqsch}\ee
The averaged potential is given by
\begin{equation}
V_m(z)=\int \!\! d^2 \vec r_{\perp}\, |W_m (\vec r_{\perp})|^2\, {1 \over r}.
\label{eqvm}\end{equation}
Here and henceforth we shall employ $\hat \rho$ as our length unit in all the 
wavefunctions and averaged potentials, 
making them dimensionless functions (thus, $f_{m\nu}''=d^2f_{m\nu}(z)/dz^2$,
with $z$ in units of $\hat\rho$). 

There are two distinct types of states in the energy spectrum $E_{m\nu}$.
The ``tightly bound'' states have no node in their $z$-wavefunctions 
($\nu=0$). The transverse size of the atom in the 
$(m,0)$ state is $L_\perp\sim\rho_m=[(2m+1)/b]^{1/2}$.
For $\rho_m\ll 1$, the atom is elongated with $L_z\gg L_\perp$.
We can estimate the longitudinal size $L_z$ by minimizing the
energy, $E\sim L_z^{-2}-L_z^{-1}\ln (L_z/L_\perp)$, giving 
\be
L_z\sim {1\over 2\ln(1/\rho_m)}=l_m^{-1},
\ee
where
\be
l_m\equiv \ln {b\over 2m+1}.
\ee
The energy of the tightly bound state is then
\be
E_m\simeq -0.16A\, l_m^2\qquad ({\rm for}~~2m+1\ll b)
\label{eqem}\ee
(recall that the energy is in units of $1\,{\rm a.u.}=e^2/a_0$).
For $\rho_m\go 1$, or $2m+1\go b$ [but still $b\gg (2m+1)^{-1}$ so that the
adiabatic approximation ($|E_m|\ll b$) is valid], 
we have $L_z\sim \rho_m^{1/2}$, and the
energy levels are approximated by
\be
E_m\simeq -0.6\,\left({b\over 2m+1}\right)^{1/2}
\qquad [{\rm for}~~2m+1\go b\gg (2m+1)^{-1}].
\label{eqem2}\ee
In Eqs.~(\ref{eqem}) and (\ref{eqem2}), the numerical coefficients
are obtained from numerical solutions of the Schr\"odinger equation
(\ref{eqsch}); the coefficient $A$ in (\ref{eqem}) is close to unity for
the range of $b$ of interest ($1\ll b\lo 10^6$) and
varies slowly with $b$ and $m$ (e.g., $A\simeq 1.01-1.3$ for $m=0-5$ when
$B_{12}=1$, and $A\simeq 1.02-1.04$ for $m=0-5$ when $B_{12}=10$.
Note that $E_m$ asymptotically approaches $-0.5\,l_m^2$ 
when $b\rightarrow\infty$;
see Hasagawa \& Howard 1961 and Haines \& Roberts 1969).
For the ground state, $(m,\nu)=(0,0)$, 
the sizes of the atomic wavefunction perpendicular and parallel to the 
field are of order $L_\perp\sim \hat\rho=b^{-1/2}$ and $L_z\sim l_0^{-1}$, 
where $l_0\equiv \ln b$. The binding energy $|E({\rm H})|$ (or the ionization
energy $Q_1$) of the atom is given by
\be
Q_1=|E({\rm H})|\simeq 0.16A\,l_0^2,
\ee
where $A$ can be approximated by
$A=1+1.36\times 10^{-2}\,[\ln (1000/b)]^{2.5}$ for $b<10^3$ and
$A=1+1.07\times 10^{-2}\,[\ln (b/1000)]^{1.6}$ for $b\ge 10^3$
(this is accurate to within $1\%$ for $100\lo b\lo 10^6$). 
Figure 1 depicts $Q_1$ as a function of $B$.
Numerical values of $Q_1$ for selected $B$'s are given in Table I.
Numerical values of $E_m$ for different $B$'s can be found, for 
example, in Ruder \etal~(1994). A fitting formula for $E_m$ 
(accurate to within $0.1-1\%$ at $0.1\lo b\lo 10^4$) 
is given by Potekhin (1998):
\be
E_m=-{1\over 2}\ln\Bigl\{\exp\left[(1+m)^{-2}\right]+p_1\left[\ln
\left(1+p_2\,b^{0.5}\right)\right]^2\Bigr\}
-{1\over 2}p_3\Bigl[\ln\left(1+p_4\,b^{p_5}\right)\Bigr]^2,
\ee
where $p_1-p_5$ are independent of $b$ (for the $m=0$ state, the parameters
$p_1-p_5$ are $15.55,\,0.378,\,2.727,\,0.3034$ and $0.438$, respectively.
Note that, unlike the field-free case, the excitation energy
$\Delta E_m=|E({\rm H})|-|E_m|$ is small compared to $|E({\rm H})|$.

Another type of states of H atom have nodes in the $z$-wavefunctions 
($\nu>0$). These states are ``weakly bound''. For example, 
the $\nu =1$ state has about the same binding energy as the ground state 
of a zero-field H atom, $E \simeq -13.6 ~{\rm eV}$, since the equation 
governing this odd-parity state is almost the same
as the radial equation satisfied by the $s$-state of the zero-field
H atom. The energy levels of the weakly bound states 
are approximately given by (Haines and Roberts 1969)
\be
E_{m\nu}=-{1\over 2\, (\nu_1+\delta)^2},\qquad (\nu_1=1,2,3,\cdots)
\label{eqmnu1}\ee
where 
\be
\delta=\cases{{2\rho_m/a_0}, & for $\nu=2\nu_1-1$;\cr
[\ln(a_0/\rho_m)]^{-1}, & for $\nu=2\nu_1$.\cr}
\label{eqmnu2}\ee
[More accurate fitting formula for $\delta$ is given in Potekhin (1998)].
The sizes of the wavefunctions are $\rho_m$
perpendicular to the field and $L_z\sim \nu^2a_0$ along the field. 

The above results assume a fixed Coulomb potential produced by the proton
(i.e., infinite proton mass). The use of a reduced electron mass
$m_em_p/(m_e+m_p)$ introduces a very small correction
to the energy [of order $(m_e/m_p)|E_{m\nu}|$]. 
However, in strong magnetic fields, the effect of the center-of-mass motion 
on the energy spectrum is complicated. An analysis of the two-body problem in
magnetic fields shows that even for the H atom ``at rest'', there is a 
proton cyclotron correction, $m\hbar\omega_{cp}=m\,(m_e/m_p)\,b$~a.u., 
to the energy [see Eq.~(\ref{eqemtot})]. We will return to this issue in
Sec.~III.E.

\begin{table}
\caption{Energy releases (in eV) in various atomic and 
molecular processes in a strong magnetic field; the values of
$Q$'s give the relative binding energies of different forms
of hydrogen in the ground state. 
Here $B_{12}=B/(10^{12}\,{\rm G})$. The zero-point
energies of the protons have been ignored in calculating
the $Q$'s for molecules. For H$_2$, the two columns give
$Q_2^{(\infty)}$ (with infinite proton mass) and $Q_2$ 
[including zero-point energy correction; see Eq.~(\ref{eqq2})];
for H$_\infty$, the two columns give $Q_\infty^{(\infty)}$ (with
infinite proton mass) and $Q_\infty$ [including zero-point 
energy correction; see Eq.~(\ref{eqqinfty})]. Note that for $B_{12}\lo 0.25$,
the lowest energy state of H$_2$ corresponds to the weakly bound state, while
for $B_{12}\go 0.25$, the tightly bound state is the ground state (see 
Sec.~IV.C). The results are
obtained using the numerical methods described in Lai \etal~(1992) and Lai
and Salpeter (1996). The numbers are generally accurate to within $10\%$.}
\begin{tabular}{c c c c c c}
  & e$+$p$=$H&H$+$e$=$H$^-$&H$+$p$=$H$_2^+$&H$+$H$=$H$_2$
& H$_{\infty}+$H$=$H$_{\infty+1}$\\ 
$B_{12}$ & $Q_1$ & $Q(\rH^-)$ & $Q(\rH_2^+)$ & $Q_2^{(\infty)}\qquad
Q_2$ & $Q_\infty^{(\infty)}\qquad Q_\infty$ \\
\hline
$0.1$ & $76.4$&$6.9$&$23.6$&$14\qquad (13)$ & $3.8\qquad (1.2)$\\
$0.5$ & $130$&$11$&$51.8$&$31\qquad (21)$  & $16.3\qquad (10)$\\
$1$ &  $161$&$13$&$70.5$&$46\qquad (32)$   & $29\qquad (20)$\\
$5$ &  $257$&$20$&$136$&$109\qquad (80)$     & $91\qquad (71)$\\
$10$& $310$&$24$&$176$&$150\qquad (110)$      & $141\qquad (113)$\\
50  & 460 & 37  & 308 & $294\qquad (236)$  & $366\qquad (306)$\\
$100$ & $541$&$42$&$380$&$378\qquad (311)$    & $520\qquad (435)$\\
$500$& $763$&$57$&$599$&$615\qquad (523)$     & $1157\qquad (964)$\\
1000 & 871  & 64 & 722 &  $740\qquad (634)$ & $1630\qquad (1350)$\\ 
\end{tabular}
\end{table}

\subsection{High-$Z$ Hydrogenic Ions}

The result in Sec.~III.A can be easily generalized to hydrogenic 
ions (with one electron and nuclear charge $Z$). The adiabatic
approximation (where the electron lies in the ground Landau level)
holds when $\hat\rho\ll a_0/Z$, or
\be
b\gg Z^2.
\ee
For a tightly bound state, $(m,\nu)=(m,0)$, the transverse size 
is $L_\perp\sim\rho_m$, while the longitudinal size is 
\be
L_z\sim \left(Z\ln {1\over Z\rho_m}\right)^{-1}.
\ee
The energy is given by
\be
E_m\simeq -0.16\,AZ^2\,\left[\ln{1\over Z^2}\left({b\over 2m+1}\right)
\right]^2
\label{eqem3}\ee
for $b\gg (2m+1)Z^2$. Results for the weakly bound states ($\nu>0$) 
can be similarly generalized from Eqs.~(\ref{eqmnu1})-(\ref{eqmnu2}).

\subsection{Heavy Atoms}

We can imagine constructing a multi-electron atom by placing electrons
at the lowest available energy levels of a hydrogenic ion. This 
picture also forms the basis for more detailed calculations of heavy atoms
which include electron-electron interactions in a self-consistent manner.

\subsubsection{Approximate Scaling Relations}

The lowest levels to be filled are the tightly bound states with $\nu=0$.
When $a_0/Z \gg \sqrt {2 Z-1} \hat\rho$, i.e., 
\be
b \gg 2 Z^3,
\label{regime1}\ee
all electrons settle into the tightly bound levels with $m=0,1,2,\cdots,Z-1$. 
The energy of the atom is approximately given
by the sum of all the eigenvalues of Eq.~(\ref{eqem3}). Accordingly, we 
obtain an asymptotic expression for for $Z \gg 1$ (Kadomtsev and Kudryavtsev
1971)
\begin{equation}
E \sim - Z^3 l_Z^2,
\ee
where
\be
l_Z=\ln \biggl ({a_0 \over Z \sqrt {2 Z-1} \hat \rho}\biggr )
\simeq \ln \sqrt {b \over 2 Z^3}.
\end{equation}
The size of the atom is given by
\be
L_\perp\sim (2Z-1)^{1/2}\hat\rho,\qquad
L_z\sim {a_0\over Zl_Z}.
\ee

For intermediate-strong fields (but still strong enough to ignore the 
Landau excitation), 
\be 
Z^{4/3} \ll b \ll 2 Z^3,
\label{regime2}\ee
many $\nu>0$ states of the inner Landau orbitals (states with relatively 
small $m$) are populated by the electrons. In this regime a Thomas-Fermi
type model for the atom is appropriate (at least for the ``core'' electrons
in small Landau orbitals), i.e., the electrons can be treated as a
one-dimensional Fermi gas in a more or less
spherical atomic cavity (Kadomtsev 1970; Mueller \etal~1971). 
The electrons occupy the ground Landau level, with the $z$-momentum up
to the Fermi momentum $p_F\sim n_e/b$, where $n_e$ is the 
number density of electrons inside the atom (recall that the degeneracy 
of a Landau level is $e B /hc \sim b$). The kinetic energy of electrons 
per unit volume is $\eps_k \sim b\,p_F^3\sim n_e^3/b^2$, and
the total kinetic energy is $E_k \sim 
R^3 n_e^3 /b^2 \sim Z^3 /b^2 R^6$, where $R$ is the radius of the atom.
The potential energy is $E_p \sim -Z^2/R$. Therefore the total energy of the
atom can be written as
\begin{equation}
E \sim {Z^3 \over b^2 R^6} - {Z^2 \over R}.
\end{equation}
Minimizing $E$ with respect to $R$ yields
\begin{equation}
R \sim Z^{1/5}b^{-2/5},\qquad E \sim -Z^{9/5} b^{2/5}.
\label{heavyatom}\end{equation}
For these relations to be valid, the electrons
must stay in the ground Landau level; this requires
$Z/R\ll\hbar\omega_{ce}=b$, which corresponds to $b\gg Z^{4/3}$.
More elaborate Thomas-Fermi type models have been developed for this regime,
giving approximately the same scaling relations (see Appendix B).  

We now consider multi-electron negative ions.
First imagine forming a H$^{-}$ ion by attaching an
extra electron to a H atom in the ground state (with $m=0$).
The extra electron can only settle into the $m=1$ state, which, 
if we ignore the screening of the proton potential due the first ($m=0$)
electron, has a binding energy of $|E_1|$ as in Eq.~(\ref{eqem}).
The Coulomb repulsion between the two electrons
reduces the binding of the $m=1$ electron. The repulsive energy is 
of order $(\ln \sqrt{b})/L_z$, which is of the same order as $|E_{1}|$. 
But the repulsive energy is smaller than $|E_{1}|$ because of the 
cylindrical charge distribution of both electrons.
Therefore  H$^{-}$ is bound relative to H$+$e and its ionization potential
is proportional to $(\ln \sqrt b)^2$. 

Similar consideration can be applied to a $Z$-ion (nuclear charge
$Z$) with $N$ electrons in the superstrong field regime ($b\gg Z^3$)
(Kadomtsev and Kudryavtsev 1971).
The sizes of the ion perpendicular and parallel to the field are,
respectively,
\begin{equation}
R\sim \sqrt{2 N-1 \over b},\qquad L_z \sim {1 \over Z l},\qquad
{\rm with}~~l = \ln \biggl ({L_z \over R} \biggr).
\end{equation}
The ground state energy of the ion is
\begin{equation}
E \simeq -{N \over 8} l^2 (4 Z-N+1)^2.
\end{equation}
Applying this result for hydrogen ions ($Z=1$), we see that the ionization
potential of H$^{-}$ to H$+$e is about $1/10$ of the binding energy of the 
H atom. Also we see that for $n > 2$, the negative ion
H$^{-(n-1)}$ is unbound. This estimate is confirmed by numerical calculations
(Lai, Salpeter and Shapiro 1992). 

\subsubsection{Numerical Calculations and Results} 

Reliable values for the 
energy of a multi-electron atom for $b\gg 1$ can be calculated
using the Hartree-Fock method (Virtamo 1976; Pr\"oschel \etal~1982;
Neuhauser \etal~1987). For an electron in the ground Landau level with spin
aligned antiparallel to the magnetic field (``adiabatic approximation''), 
the kinetic energy and spin energy add up to $(1/2 m_e) p_{z}^2$. Thus the
Hamiltonian for a neutral atom with $Z$ electrons is 
\be
H =\sum_{i}{1 \over 2 m_e} p_{zi}^2
-Ze^2 \sum_i {1 \over r_i}+e^2 \sum_{i<j} {1 \over r_{ij}},
\ee
where $i$ labels electrons. The basis one-electron wavefunctions are  
\begin{equation}
\Phi _{m \nu}(\br)=W_m (\rho,\phi) f_{m \nu}(z),\qquad m,\nu=0,1,2,\cdots
\end{equation}
In the Hartree-Fock approximation, the $Z$-electron wave function is formed by 
the antisymmetrized product of $Z$ one-electron basis functions. 
Averaging the Hamiltonian over the transverse Landau functions, we obtain
\begin{eqnarray}
\langle H \rangle &=& {\hbar^2 \over 2 m_e {\hat \rho}^2} \sum_{m\nu}
\int \! dz\,|f'_{m,\nu}(z)|^2
-{Ze^2 \over \hat \rho}\sum_{m\nu}\int \! dz\, |f_{m,\nu}(z)|^2 V_m(z)
\nonumber \\
&&+E_{\rm dir}+E_{\rm exch},
\ea
where $V_m(z)$ is given by Eq.~(\ref{eqvm}). The direct and exchange energies
for electron-electron interaction are given by
\ba
E_{\rm dir}&=& {e^2 \over 2 \hat \rho}\sum_{m\nu,m'\nu'}\int\!dz dz'\,
D_{mm'}(z-z')|f_{m\nu}(z)|^2 |f_{m'\nu'}(z')|^2,\\
E_{\rm exch}&=& -{e^2 \over 2 \hat \rho}\sum_{m\nu,m'\nu'}\int\!dz dz'\,
E_{mm'}(z-z')f_{m\nu}(z) f_{m'\nu'}(z') f^*_{m\nu}(z')f^*_{m'\nu'}(z).
\end{eqnarray}
where 
\begin{eqnarray}
D_{mm'} (z_1-z_2) &=& \int \! d^2 \br_{1 \perp} d^2 \br_{2 \perp}
|W_m(\br_{1 \perp})|^2 |W_{m'}(\br_{2 \perp})|^2\, {1 \over r_{12}},\\
E_{mm'} (z_1-z_2) &=& \int \! d^2 \br_{1 \perp} d^2 \br_{2 \perp}
W_m(\br_{1 \perp}) W_{m'}(\br_{2 \perp}) 
W_m^*(\br_{2 \perp}) W_{m'}^*(\br_{1 \perp})\,{1 \over r_{12}}.
\label{eqemm}\end{eqnarray}
(Useful mathematical relations for evaluating $V_m, D_{mm'}, E_{mm'}$
are given in, e.g., Sokolov and Ternov 1968; Virtamo and Jauho 1975;
Pr\"oschel \etal~1982; Lai \etal~1992).
Varying $\langle H \rangle$ with respect to $f_{m\nu}$'s, 
we obtain the Hartree-Fock equations
\begin{equation}
\biggl [-{ \hbar^2 \over 2 m_e {\hat \rho}^2}
{d^2 \over dz^2} - {Ze^2 \over \hat \rho} V_m(z)
+ {e^2 \over \hat \rho} K_m (z) - \eps_{m\nu} \biggr] f_{m\nu}(z)=
{e^2 \over \hat \rho} J_{m \nu}(z),
\label{eqhf}\end{equation}
where the direct and exchange potentials are
\begin{eqnarray}
K_m(z) &=& \sum_{m'\nu'}\int\!dz' |f_{m'\nu'}(z')|^2 D_{mm'}(z-z'), \\
J_{m\nu}(z) &=& \sum_{m'\nu'} f_{m'\nu'}(z) \int\!dz'\, f^*_{m'\nu'}(z') 
f_{m\nu}(z') E_{mm'}(z-z').
\label{eqjm}\end{eqnarray}
After iteratively solving Eqs.~(\ref{eqhf})-(\ref{eqjm}) for 
the eigenvalues $\varepsilon_{m\nu}$ and eigenfunction $f_{m\nu}$, 
the total energy of the atom can be obtained from
\be
E=\sum_{m\nu}\varepsilon_{m\nu}-E_{\rm dir}-E_{\rm exch}.
\label{eqehf}\ee

Accurate energies of He atom as a function of $B$ in the adiabatic 
approximation (valid for $b\gg Z^2$) were obtained by Virtamo (1976) 
and Pr\"oschel \etal~(1982). This was extended to $Z$ up to $26$ (Fe atom) by 
Neuhauser \etal~(1987) (see also Miller and Neuhauser 1991).
Numerical results can be found in these papers. 
Neuhauser \etal~(1987) gave an approximate fitting formula 
\be
E\simeq -160\,Z^{9/5}B_{12}^{2/5}\,{\rm eV}
\label{eqeatom}\ee
for $0.5\lo B_{12}\lo 5$. (Comparing with the numerical results, the accuracy 
of the formula is about $1\%$ for $Z\simeq 18-26$ and
becomes $5\%$ for $Z\sim 10$.) For the He atom, more accurate results
(which relax the adiabatic approximation) are given in
Ruder \etal~(1994) and in Jones \etal~(1999a) (this paper also considers 
the effect of electron correlation).   

The Hartree-Fock method is approximate because electron correlations are
neglected. Due to their mutual repulsion, any pair of electrons tend to be
more distant from each other than the Hartree-Fock wave function would 
indicate. In zero-field, this correlation effect is especially
pronounced for the spin-singlet states of electrons for which the spatial 
wave function is symmetrical. In strong magnetic fields, the electron spins are
all aligned antiparallel to the magnetic field, the spatial wavefunction is
antisymmetric with respect to the interchange of two electrons. 
Thus the error in the Hartree-Fock approach is expected to be 
significantly smaller than the $1\%$ accuracy characteristic of zero-field
Hartree-Fock calculations (Weissbluth 1978; Neuhauser \etal~1987;
Schmelcher, Ivanov and Becken 1999). 

Other calculations of heavy atoms in strong magnetic fields include
Thomas-Fermi type statistical models (see Fushiki \etal~1992
for a review and Appendix B for a brief summary) 
and density functional theory (Jones 1985,~1986;
K\"ossl \etal~1988; Relovsky and Ruder 1996). 
The Thomas-Fermi type models are useful in establishing asymptotic scaling
relations, but are not adequate for obtaining accurate binding energy
and excitation energies. The density functional theory can potentially 
give results as accurate as the Hartree-Fock method, but, without 
calibration with other methods, it is difficult to establish its 
accuracy {\it a priori} (see Neuhauser \etal~1987; Vignale and Rasolt
1987,1988). Lieb \etal~(1992,1994a,b) have presented detailed
discussions of the asymptotic behaviors of heavy atoms (as
$Z\rightarrow\infty$) for five different magnetic field regimes: $b\ll
Z^{4/3}$, $b\sim Z^{4/3}$, $Z^{4/3}\ll b\ll Z^3$, $b\sim Z^3$ and $b\gg Z^3$
[see also eqs.~(\ref{regime1})-(\ref{regime2})]. They showed that various
density functional type theories become exact in these asymptotic limits: 
the Thomas-Fermi theory corresponds to the first three regimes, and 
a ``density-matrix functional'' theory can be applied to the fifth regime
[see Johnsen \& Yngvason (1996) for numerical calculations of heavy atoms 
based on this theory].

\subsection{Intermediate Magnetic Field Regime}

For $B\sim B_0\sim 10^9$~G, the adiabatic approximation is no longer valid,
and electrons can occupy excited Landau levels. In this intermediate field
regime, neither Coulomb nor magnetic effects can be treated as a perturbation.
Accurate energy levels of H atom for arbitrary field strengths were first
calculated by R\"osner \etal~(1984); the method involves expansion
of the wavefunction in terms of either spherical harmonics or 
cylindrical Landau orbitals, and subsequent approximate solution of the system 
of the coupled integral-differential equations (See Ruder \etal~1994 for
tabulated numerical results). Recent calculations of H atom in 
magnetic fields include Goldman and Chen (1991),
Chen and Goldman (1992) (relativistic 
effects), Melezhik (1993), Fassbinder and Schweizer (1996) (with magnetic
and electric fields), Kravchenko \etal~(1996) (exact solution for
nonrelativistic electron) and references therein. Ruder \etal~(1994) presented
(less accurate) results for He atoms calculated using a similar Hartree-Fock
method as in R\"osner \etal~(1984). Recent studies of multi-electron atoms 
(including radiative transitions) for intermediate field regime have used more
elaborate implementations of Hartree-Fock method with different basis functions
(see Jones \etal~1996,~1999a, Becken \etal~1999,
Becken and Schmelcher 2000, Ivanov and Schmelcher 1998,1999,
2000). Accurate calculation of the hydrogen negative ion (H$^-$) was 
presented by Al-Hujaj and Schmelcher (2000) (and references therein). 
Quantum Monte Carlo method has also been
developed to calculate He atom (Jones \etal~1997).
Accurate energy levels for He atom at $B\sim 10^9$~G are needed to interpret
the spectrum of magnetic white dwarf GD229 (see Jordan \etal~1998;
Jones \etal~1999b).

\subsection{Effect of Center-of-Mass Motion}

Our discussion so far has implicitly assumed infinite nuclear mass,
i.e., we have been concerned with the energy levels of electrons
in the static Coulomb potential of a fixed ion. It has long been
recognized that in strong magnetic fields, the effects of finite nuclear mass
and center-of-mass (CM) motion on the atomic structure are nontrivial
(e.g., Gor'kov and Dzyaloshinskii 1968; Avron \etal~1978; Herold \etal~1981;
Baye and Vincke 1990; Vincke \etal~1992; Pavlov and Meszaros 1993; Potekhin
1994). Here we illustrate 
the key issues by considering the hydrogen atom in strong magnetic fields: 
general quantum mechanical solutions for this two-body problem 
have been obtained only recently (Vincke \etal~1992; Potekhin 1994).
Some of the aspects are also important for applications to molecules.
A recent discussion on this subject can be found in Baye and Vincke (1998) 
(see also Johnson \etal~1983 for an earlier review on the general problem of
CM motion of atoms and molecules in external fields). 

A free electron confined to the ground Landau level, the usual case for 
$b\gg 1$, does not move perpendicular to the magnetic field. Such motion
is necessarily accompanied by Landau excitations. 
When the electron combines with a proton the mobility of the neutral 
atom across the field depends on the ratio of the atomic excitation energy 
($\sim \ln b$) and the Landau excitation energy for the proton,
$\hbar\omega_{cp}=\hbar eB/(m_pc)$. It is convenient to 
define a critical field strength $B_{\rm cm}$ via
(Lai and Salpeter 1995).
\be
b_{\rm cm}\equiv {m_p\over m_e}\ln b_{\rm cm}=1.80\times 10^4;\qquad
B_{\rm cm}=b_{\rm cm}B_0=4.23\times 10^{13}~{\rm G}.
\label{bcrit}\ee
Thus for $B\go B_{\rm cm}$, the deviation from the free 
center-of-mass motion of the atom is significant even for
small transverse momentum [see Eq.~(\ref{mperp}) below].

Consider now the electron-proton system. It is easy to show that even 
including the Coulomb interaction, the total pseudomomentum,
\be
\bK=\bK_e+\bK_p,
\ee
is a constant of motion, Moreover, all components of $\bK$ commute with each
other. Thus it is natural to separate the CM motion from the internal degrees
of freedom using $\bK$ as an explicit constant of motion. From 
Eq.~(\ref{eqbrc}), we find that the separation between the guiding 
centers of the electron and the proton is directly related to $\bK_\perp$:
\be
\bR_K=\bR_{ce}-\bR_{cp}={c\bB\times \bK\over eB^2}.
\ee
The two-body eigenfunction with a definite value of $\bK$ can be written as
\be
\Psi(\bR,\br)=\exp \left[{i\over\hbar}\left(\bK+{e\over 2c}
\bB\times\br\right)\cdot\bR\right]\phi(\br),
\ee
where $\bR=(m_e\br_e+m_p\br_p)/M$ and $\br=\br_e-\br_p$ are the center-of-mass
and relative coordinates, and $M=m_e+m_p$ is the total mass. The
Schr\"odinger equation reduces to
$H\phi(\br)=\cE\,\phi(\br)$, with
\footnote{The spin terms of the electron and
the proton are not explictly included. For the electron 
in the ground Landau level, the zero-point Landau energy 
$\hbar\omega_{ce}/2$ is exactly cancelled by the spin energy. 
For sake of brevity, we drop the zero-point energy of the proton,
$\hbar\omega_{cp}/2$, as well as the spin energy $\pm
g_p\hbar\omega_{cp}/2$ (where $g_p=2.79$).}
\be
H={\bK^2\over 2M}+{1\over 2\mu}\left(\bp+{e\over 2c}\bB\times\br\right)^2
-{e\over m_p c}\bB\cdot (\br\times\bp)-{e^2\over r}
+{e\over Mc}(\bK\times\bB)\cdot\br,\label{eqha1}
\ee
where $\bp=-i\hbar\partial/\partial\br$ and $\mu=m_em_p/M$
(e.g., Lamb 1952; Avron \etal~1978; Herold \etal~1981). 
Clearly, the CM motion is coupled to the internal motion 
through the last term in $H$, which has the form of an electrostatic 
potential produced by an electric field $(\bK/M)\times\bB$.
This term represents the so called ``motional Stark effect''
(although such a description is not exactly accurate, since
$\bK_{\perp}/M$ does not correspond to the CM velocity; see Johnson 
\etal~1983). In the adiabatic approximation ($b\gg 1$, with the electron
in the ground Landau level), we write the total energy of the atom as
\be
\cE_{m\nu}(K_z,K_\perp)={K_z^2\over 2M}+m\hbar\omega_{cp}+E_{m\nu}(K_\perp),
\label{eqemtot}\ee
where azimuthal quantum number $m$ measures the relative 
electron-proton $z$-angular momentum $J_z=\hat\bB\cdot (\br\times\bp)=
-m$ (clearly, $m$ is a
good quantum number only when $K_\perp=0$, but we shall use it as a label
of the state even when $K_\perp\ne 0$), and $\nu$ enumerates the longitudinal
excitations. The $m\hbar\omega_{cp}$ term in Eq.~(\ref{eqemtot}) represents 
the Landau energy excitations for the proton; this ``coupling'' between the
electron quantum number $m$ and the proton Landau excitation results from the
conservation of $\bK$. Clearly, for sufficiently high $b$, states 
with $m>0$ become unbound [i.e.,
$\cE_{m\nu}(\bK=0)=m\hbar\omega_{cp}+E_{m\nu}(0)>0$]. 

For small $K_\perp$, the motional Stark term in $H$ can be treated 
as a perturbation (e.g., Vincke and Baye 1988; Pavlov and Meszaros 1993). 
For the tightly bound states ($\nu=0$), we have
\be
E_{m0}(K_\perp)\simeq E_m+{K_\perp^2\over 2M_{\perp m}},
\qquad ({\rm for}~~K_\perp\ll K_{\perp p})
\label{eqemtot1}
\ee
where $E_{m}$ is the energy of a bound electron in the fixed Coulomb potential
(the correction due to the reduced mass $\mu$ can be easily incorporated 
into $E_m$; this amounts to a small correction). The effective mass 
$M_{\perp m}$ for the transverse motion increases with increasing $b$. For
the $m=0$ state, 
\be
M_{\perp}\equiv M_{\perp 0}= M\left(1+{\xi b\over M\ln b}\right)
\simeq M\left(1+{\xi b\over b_{\rm cm}}\right),
\label{mperp}\ee
where $\xi$ is a slowly varying function of $b$ (e.g., $\xi\simeq 2-3$
for $b=10^2-10^5$). Similar results can be obtained for the $m>0$ states:
$M_{\perp m}\simeq M+\xi_m (2m+1){b/ l_m}$, where $\xi_m$ is of the same order
of magnitude as $\xi$, and $l_m=\ln[b/(2m+1)]$. A simple fitting
formula for $M_{\perp m}$ is proposed by Potekhin (1998):
\be
M_{\perp m}=M\left[1+(b/b_0)^{c_0}\right],
\ee
with $b_0=6150\,(1+0.0389\,m^{3/2})/(1+7.87\,m^{3/2})$ and
$c_0=0.937+0.038\,m^{1.58}$.
Equation (\ref{eqemtot1}) is valid only when $K_\perp$ is much less
than the ``perturbation limit'', $K_{\perp p}$, given by (for $m=0$)
$K_{\perp p}=b^{1/2}\left(1+{M\ln b/\xi b}\right)$,
which corresponds to $K_{\perp p}^2/(2M_\perp)\simeq 1.7
\left(1+{b_{\rm cm}/\xi b}\right)$. Numerical calculations by Potekhin (1994) 
indicate that Eq.~(\ref{eqemtot1}) is a good approximation for
$K_\perp\lo K_{\perp c}$ (see Fig.~2), where
\be
K_{\perp c}\simeq \sqrt{2M|E_m|},
\ee
that is,  $|E_{m0}(K_{\perp c})|\ll |E_m|$. The states with $K_\perp\lo
K_{\perp c}$ are sometimes called ``centered states''. 

For $K_\perp\go K_{\perp c}$, the motional electric field [see 
Eq.~(\ref{eqha1})] induces a significant transverse 
separation (perpendicular to $\bK$) between the electron and the proton. 
It is more convenient to use a new coordinate system to account
for this effect. Since $\bK_{\perp}$ measures the
separation of the guiding centers of the electron and the proton, we can 
remove the ``Stark term'' in Eq.~(\ref{eqha1}) 
by introducing a displaced coordinate
$\br'=\br-\bR_K$. After a gauge transformation with
\be
\phi(\br)\rightarrow 
\exp\left({i\over\hbar}\,{m_p-m_e\over 2M}\bK_\perp\cdot\br\right)\phi(\br'),
\ee
we obtain $H'\phi(\br')=\cE\phi(\br')$, with the Hamiltonian 
\be
H'={K_z^2\over 2M}+{1\over 2\mu}\left(\bp'+{e\over 2c}\bB\times\br'\right)^2
-{e\over m_pc}\bB\cdot (\br'\times\bp')-{e^2\over |\br'+\bR_K|},
\label{eqha3}\ee
where $\bp'=-i\hbar\partial/\partial\br'$ (e.g., Avron \etal~1978; Herold et
al.~1981). We can estimate the size $L_z$ of the atom along the $z$-axis and
the energy $E_{m\nu}(K_\perp)$ for two different regimes of $R_K$
(see Lai and Salpeter 1995): (i) For
$R_K\lo L_z\lo 1$ (but not necessarily $R_K<\rho_m$ or $K_\perp<
\sqrt{mb}$), we have (for the $\nu=0$ states)
\be
L_z\sim \left(\ln {1\over\rho_m^2+R_K^2}\right)^{-1},\qquad
E_{m0}(K_\perp)\sim -0.16\,\left(\ln {1\over\rho_m^2+R_K^2}\right)^2.
\ee
Mixing between different $m$-states is unimportant when 
$b\gg b_{\rm cm}$. (ii) For $R_K\go 1$, the electron-proton interaction does
not contain the Coulomb logarithm, and the energy can be written as
$E_{m0}(K_\perp)\sim L_z^{-2}-(L_z^2+R_K^2)^{-1/2}$.
In the limit of $R_K\gg 1$, minimization of $\cE_m$ with respect to $L_z$
yields 
\be
L_z\sim R_K^{3/4}\ll R_K,\qquad
E_{m0}(K_\perp)\sim -{1\over R_K}=-{b\over K_\perp},
\ee
independent of $m$ 
(e.g., Burkova \etal~1976; Potekhin 1994). The states with $K_\perp\go
K_{\perp c}$ (which corresponds to $R_K\go\hat\rho$ for $b\go b_{\rm cm}$) are
sometimes referred to as ``decentered states'' (Vincke \etal~1992; Potekhin
1994), but note that for a given $(m\nu)$ (defined at $K_\perp=0$), there is a
continuous energy $E_{m\nu}(K_\perp)$ which connects the ``centered
state'' at small $K_\perp$ to the ``decentered state'' at large $K_\perp$. 

To calculate the energy $E_{m\nu}(K_\perp)$ for general $K_\perp$, 
we must include mixing between different $m$-orbitals. We may use
$\phi(\br)=\sum_{m'}W_{m'}(\br_\perp)f_{m'}(z)$ in 
Eq.~(\ref{eqha1}) and obtain a coupled set of equations for
$f_{m'}(z)$; alternatively, for $K_\perp\go K_{\perp c}$, it is more
convenient to use $\phi(\br')=\sum_{m'}W_{m'}(\br_\perp')f_{m'}(z)$
in Eq.~(\ref{eqha3}). Numerical results are presented in Potekhin (1994)
(see also Vincke \etal~1992); analytical fitting formulae for the energies,
atomic sizes and oscillator strengths are given in Potekhin (1998). 
Figure 2 shows the energy $E_{m\nu}(K_\perp)$ (based on Potekhin's
calculation) as a function of $K_\perp$ for several different states at
$B_{12}=1$. Note that the total energies of different states, $\cE_{m\nu}(K_z,
K_\perp)$, do not cross each other as $K_\perp$ increases. 

While for neutral atoms the CM motion can be separated from the internal
relative motion, this cannot be done for ions (Avron \etal~1981).
Ions undergo collective cyclotron motion which depends on the internal state. 
However, the existence of an approximate constant of motion allows
an approximate pseudoseparation up to very high fields (see Bayes and Vincke
1998 and references therein). Numerical results for He$^+$ moving in strong
magnetic fields are obtained by Bezchastnov \etal~(1998). 

The effects of CM motion on multi-electron systems (heavy atoms and molecules)
in strong magnetic fields have not been studied numerically, although 
many theoretical issues are discussed in Johnson \etal~(1983) and 
Schmelcher \etal~(1988,1994).

\section{Molecules}

Most studies of molecules in strong magnetic fields have been 
restricted to hydrogen. We will therefore focus on H to 
illustrate the basic magnetic effects and only briefly discuss
molecules of heavier elements.

\subsection{H$_2$: Basic Mechanism of Bonding}

In a strong magnetic field, the mechanism of forming
molecules is quite different from the zero-field case
(see Ruderman 1974; Lai \etal~1992). The spins of the electrons in the atoms
are aligned anti-parallel to the magnetic field,
and therefore two atoms in their ground states ($m=0$) 
do not bind together according to the exclusion principle. 
Instead, one H atom has to be excited to the $m=1$ state. The
two H atoms, one in the ground state ($m=0$), another
in the $m=1$ state then form the ground state of the H$_2$ molecule by
covalent bonding. Since the ``activation energy'' for exciting
an electron in the H atom from the Landau orbital $m$ to $(m+1)$
is small [see Eq.~(\ref{eqem})], the resulting H$_2$ molecule is stable.
The size of the H$_2$ molecule is comparable to that of the H atom. 
The interatomic separation $a_{\rm eq}$ and the dissociation energy $D$ of the
H$_2$ molecule scale approximately as 
\begin{equation}
a_{\rm eq}\sim {1\over \ln b},\qquad D\sim (\ln b)^2,
\end{equation}
although $D$ is numerically smaller than the ionization energy of the H atom
(see Table I).

Another mechanism of forming a H$_2$ molecule in a strong magnetic 
field is to let both electrons occupy the same $m=0$ Landau orbital,
while one of them occupies the tightly bound
$\nu=0$ state and the other the $\nu=1$ weakly bound state.
This costs no ``activation energy''. However, the resulting molecule 
tends to have a small dissociation energy, of order a Rydberg.
We shall refer to this electronic state of the molecule as the 
weakly bound state, and to the states formed by two electrons
in the $\nu=0$ orbitals as the tightly bound states.
As long as $\ln b\gg 1$, the weakly bound states 
constitute excited energy levels of the molecule.

\subsection{Numerical Calculations and Results}

In the Born-Oppenheimer approximation (see Schmelcher \etal~1988,1994 for a 
discussion on the validity of this approximation in strong magnetic fields),
the interatomic potential $U(a,R_{\perp})$ is given by the total 
electronic energy $E(a,R_{\perp})$ of the system, where
$a$ is the proton separation along the magnetic field, and
$R_{\perp}$ is the separation perpendicular to the field.
Once $E(a,R_{\perp})$ is obtained, the electronic equilibrium state
is determined by locating the minimum of the $E(a,R_\perp)$ surface.
[For a given $a$, $E(a,R_\perp)$ is minimal at $R_\perp=0$].

The simplest system is the molecular ion H$_2^+$.
For $b\gg 1$, the energy of H$_2^+$ can be 
easily calculated as in the case of H atom (e.g., Wunner \etal~1982;
Le Guillou and Zinn-Justin 1984; Khersonskii 1984). 
When the molecular axis is aligned with 
the magnetic axis ($R_\perp=0$), we only need to solve the Schr\"odinger 
equation similar to (\ref{eqsch}), except replacing $V_m(z)$ by
\be
\tilde V_m(z)=V_m\biggl(z-{a\over 2}\biggr)+V_m\biggl(z+{a\over 2}\biggr).
\ee
The total electronic energy is simply
$E(a,0)=\eps_{m\nu}+{e^2/a}$. The ground state corresponds to
$(m,\nu)=(0,0)$.

The Hartree-Fock calculation of H$_2$ is similar to the case of multi-electron
atoms. For the tightly bound states, the two electrons occupy the
$(m,\nu)=(m_1,0),(m_2,0)$ orbitals (the ground state corresponds to
$m_1=0$, $m_2=1$), with the wavefunction 
\be
\Psi (\br_1,\br_2)={1 \over \sqrt 2}
[\Phi_{m_10}(\br_1)\Phi_{m_20}(\br_2)
-\Phi_{m_10}(\br_2)\Phi_{m_20}(\br_1)].
\label{eqpsi1}\ee
We obtain the same Hartree-Fock equation as (\ref{eqhf})
except that $V_m$ is replaced by $\tilde V_m$ (for aligned configurations,
$R_\perp=0$). The energy is given by
\begin{eqnarray}
E (a,0)&=&{e^2\over a}+\eps_{m_10}+\eps_{m_20}-
{e^2\over\hat\rho}\int\!\! dz_1dz_2|f_{m_10}(z_1)|^2|f_{m_20}(z_2)|^2
D_{m_1m_2}(z_1-z_2)\nonumber\\
&+& {e^2\over\hat\rho}
\int\!\! dz_1dz_2f_{m_10}(z_1)f_{m_20}(z_2)f^*_{m_10}(z_2)f^*_{m_20}(z_1)
E_{m_1m_2}(z_1-z_2).
\end{eqnarray}

Although the Hartree-Fock method is adequate for small interatomic 
separations ($a$ less than the equilibrium value, $a_{\rm eq}$),
the resulting $E(a,0)$ becomes less reliable for large $a$: 
as $a\rightarrow\infty$, $E(a,0)$ does not approach the sum of the 
energies of two isolated atoms, one in the $m_1$-state, another in 
the $m_2$-state. The reason is that as $a$ increases, 
another configuration of electron orbitals,
\begin{equation}
\Psi_2 (\br_1,\br_2)={1 \over \sqrt 2}
[\Phi_{m_11}(\br_1)\Phi_{m_21}(\br_2)
-\Phi_{m_11}(\br_2)\Phi_{m_21}(\br_1)],\end{equation}
becomes more and more degenerate with the first configuration 
$\Psi_1=\Psi$ in Eq.~(\ref{eqpsi1}), 
and there must be mixing of these two different configurations.
Both $\Psi_1$ and $\Psi_2$ have the 
same symmetry with respect to the Hamiltonian:
the total angular momentum along the $z$-axis is $M_{Lz}=1$, the 
total electron spin is $M_{Sz}=-1$, and both $\Psi_1$ and 
$\Psi_2$ are even with respect to the operation $\br_i\rightarrow -\br_i$. 
To obtain a reliable $E(a,0)$ curve, ``configuration interaction'' 
between $\Psi_1$ and $\Psi_2$ must be taken into account in the
Hartree-Fock scheme (Lai \etal~1992; see, e.g., Slater 1963 for 
a discussion of the zero-field case). 

Molecular configurations with $R_\perp\neq  0$ correspond to excited states of
the molecules (see Sec.~IV.C). To obtain $E(a,R_\perp)$, 
mixing of different $m$-states in single-electron orbital
needs to be taken into account. Approximate energy surfaces 
$E(a,R_\perp)$ for both small $R_\perp$ and large $R_\perp$
have been computed by Lai and Salpeter (1996).

Numerical results of $E(a,0)$ (based on the Hartree-Fock
method) for both tightly bound states and weakly bound states are given in 
Lai \etal~(1992) and Lai and Salpeter (1996). Quantum Monte Carlo calculations
have also been performed, confirming the validity of the 
method (Ortiz \etal~1995). Figure 3 depicts some of the energy curves. The
dissociation energy of H$_2$ in the ground state can be fitted by
\be
Q_2^{(\infty)}\equiv 2|E(\rH)|-|E(\rH_2)|
=0.106\,\left[1+0.1\,l^{0.2}\ln\left(b/b_{\rm cm}\right)\right]\,
l^2,
\label{q2infty}\ee
(where $l=\ln b$), with an accuracy of $\lo 5\%$ for $1\lo B_{12}\lo 1000$, 
where $b_{\rm cm}=1.80\times 10^4$ is defined in Eq.~(\ref{bcrit})
(The superscript ``$(\infty)$'' implies that the zero-point energy 
of the molecule is not included in $Q_2^{(\infty)}$; see Sec.~IV.C below).
Thus $Q_2^{(\infty)}\simeq 46$ eV for $B_{12}=1$ and 
$Q_2^{(\infty)}\simeq 150$ eV for $B_{12}=10$ (see Table I).
By contrast, the zero-field dissociation energy of H$_2$ is $4.75$ eV.

\subsection{Molecular Excitations}

For the ground state of H$_2$, the molecular axis and the magnetic field 
axis coincide, and the two electrons occupy the $m=0$ and $m=1$ orbitals, 
i.e., $(m_1,m_2)=(0,1)$. The molecule can have different types of 
excitation levels (Lai and Salpeter 1996):

(i) {\it Electronic excitations}.
The electrons occupy orbitals other than $(m_1,m_2)=(0,1)$, 
giving rise to the electronic excitations. The energy difference 
between the excited state $(m_1,m_2)$ (with $\nu_1=\nu_2=0$) 
and the ground state $(0,1)$ is of order $\ln b$, 
as in the case for atoms. Typically, only the single-excitation levels 
(those with $m_1=0$ and $m_2>1$) are bound relative to two atoms in the 
ground states. Another type of electronic excitation is formed by two electrons
in the $(m,\nu)=(0,0)$ and $(0,1)$ orbitals.  The dissociation energy of
this weakly bound state is of order a Rydberg, and does not depend
sensitively on the magnetic field strength (see Fig.~3).
Note that since it does not cost any activation energy to form a H$_2$ in
the weakly bound state, for relatively small magnetic field
($B_{12}\lo 0.25$), the weakly bound
state actually has lower energy than the tightly bound state.

(ii) {\it Aligned vibrational excitations}.
These result from the vibration of the protons about the
equilibrium separation $a_{\rm eq}$ along the magnetic field axis.
To estimate the excitation energy, we need to consider the
excess potential $\delta U(\delta a)=U(a_{\rm eq}+\delta a,0)-U(a_{\rm eq},0)$.
Since $a_{\rm eq}$ is the equilibrium position, the sum of the 
first order terms (proportional to $\delta a$) in $\delta U$,
coming from proton-proton, electron-electron,
proton-electron Coulomb energies and quantum mechanical electron
kinetic energy, must cancel, and $\delta U\propto (\delta a)^2$ for small
$\delta a$. The dominant contribution to the energy of the molecule
comes from the proton-electron Coulomb energy
$\sim l/a$, where the logarithmic factor $l\equiv\ln b\gg 1$ results from 
the Coulomb integral over the elongated electron distribution. 
The excess potential is of order $\delta U
\sim l\,(\delta a)^2/a_{\rm eq}^3\sim \left(\xi^{-3}l^4\right)\,(\delta a)^2$,
where we have used $a_{\rm eq}=\xi/l$ (the dimensionless factor $\xi$ 
decreases slowly with increasing $b$; e.g., $\xi\simeq 2$ for $B_{12}=0.1$ and
$\xi=0.75$ for $B_{12}=100$). Thus for small-amplitude oscillations 
around $a_{\rm eq}$, we obtain a harmonic oscillation spectrum 
with excitation energy quanta $\hbar \omega_{\parallel} \sim
\xi^{-3/2}\,l^2\,\mu^{-1/2}$, where $\mu=m_p/2m_e=918$ is the reduced 
mass of the two protons in units of the electron 
mass. Numerical calculations yield a similar scaling relation.
For the electronic ground state, the energy quanta can be approximated by
\be
\hbar\omega_\parallel\simeq 0.13\,(\ln b)^{5/2}\mu^{-1/2} ({\rm a.u.})
\simeq 0.12\,(\ln b)^{5/2}~({\rm eV}).
\ee
(This is accurate to within $10\%$ for $40\lo b\lo 10^4$).
Thus $\hbar\omega_{\parallel}\simeq 10$ eV at $B_{12}=1$ 
and $\hbar\omega_{\parallel}\simeq 23$ eV at $B_{12}=10$, in contrast to 
the vibrational energy quanta $\hbar\omega_{\rm vib}\simeq 0.52$ eV for 
the H$_2$ molecule at zero magnetic field. 

(iii) {\it Transverse vibrational excitations}.
The molecular axis can deviate from the magnetic field direction,
precessing and vibrating around the magnetic axis.
Such an oscillation is the high-field analogy of
the usual molecular rotation; the difference is that in strong magnetic 
fields, this ``rotation'' is constrained around the magnetic field line. 
To obtain the excitation energy, we need to estimate the excess potential 
$\delta U(R_\perp)\equiv U(a_{\rm eq},R_\perp)-U(a_{\rm eq},0)$.
When the protons are displaced by $\sim R_\perp$ from the 
electron distribution axis, the proton-electron interaction 
is approximately given by $a_{\rm eq}^{-1}\ln \left[L_z/(\hat\rho^2+R_\perp^2)^{1/2}
\right]$. Thus an order of magnitude expression for $\delta U$ is 
$\delta U(R_{\perp})\sim (1/2a_{\rm eq})\ln\left(1+\hat\rho^{-2}R_\perp^2\right)
\sim \xi^{-1}l\,\ln\left(1+b\,R_\perp^2\right)$.
This holds for any $R_\perp\ll a_{\rm eq}=\xi\,l^{-1}$. For 
small-amplitude transverse oscillations, 
with $R_\perp\lo \hat\rho=b^{-1/2}\ll a_{\rm eq}$,
we have $\delta U\sim \xi^{-1}l\,b\,R_\perp^2$. The energy quanta is then
$\hbar \omega_{\perp 0}\sim \left(\xi^{-1}\,l\,b\right)^{1/2}\mu^{-1/2}$,
where the subscript $0$ indicates that we are at the moment neglecting 
the magnetic forces on the protons which, in the absence of Coulomb
forces, lead to proton cyclotron motions (see below). 
Numerical calculations give a similar scaling relation.
For the electronic ground state, the excitation energy quanta 
$\hbar\omega_{\perp 0}$ can be approximated by
\be
\hbar\omega_{\perp 0}\simeq 0.125\,b^{1/2}(\ln b)\mu^{-1/2} ({\rm a.u.})
\simeq 0.11\,b^{1/2}(\ln b)~({\rm eV}).
\label{homega0}\ee
(This is accurate to within $10\%$ for $40\lo b\lo 10^4$).
Thus $\hbar\omega_{\perp 0}\simeq 14$ eV at $B_{12}=1$ and 
$\hbar\omega_{\perp 0}=65$ eV at $B_{12}=10$.
(See Lai and Salpeter 1996 for a discussion of large-amplitude
oscillations.)

Note that in a strong magnetic field,
the electronic and (aligned and transverse) vibrational 
excitations are all comparable, with
$\hbar\omega_{\perp 0}\go\hbar\omega_\parallel$. This is in contrast 
to the zero-field case, where we have
$\Delta \eps_{\rm elec} \gg \hbar \omega_{\rm vib}
\gg \hbar \omega_{\rm rot}$.

Equation (\ref{homega0}) for the zero-point energy of the transverse
oscillation includes only the contribution of the electronic restoring
potential $\mu\omega_{\perp 0}^2R_\perp^2/2$. Since the magnetic forces on the
protons also induce a ``magnetic restoring potential''
$\mu\omega_{cp}^2R_\perp^2/2$, where $\hbar\omega_{cp}=\hbar eB/(m_pc)
\simeq 6.3\,B_{12}$ eV is the cyclotron energy of the proton,
the zero-point energy of the transverse oscillation is
\be
\hbar\omega_\perp=\hbar (\omega_{\perp 0}^2+\omega_{cp}^2)^{1/2}
-\hbar\omega_{cp}.
\ee
The dissociation energy of H$_2$, taking into account the zero-point 
energies of aligned and transverse vibrations, is then
\be
Q_2= Q_2^{(\infty)}-\left({1\over 2}\hbar\omega_\parallel
+\hbar \omega_\perp\right).
\label{eqq2}\ee
Some numerical values are given in Table I. Variation of $Q_2$ 
as a function of $B$ is depicted in Fig.~1.

\subsection{H$_N$ Molecules: Saturation}

At zero magnetic field, two H atoms in their ground states with spins 
opposite to each other form a H$_2$ molecule by covalent bonding;
adding more H atoms is not possible by the exclusion principle (unless
one excites the third atom to an excited state; but the resulting 
H$_3$ is short-lived). In a strong magnetic field,
the spins of the electrons in the atoms are all aligned anti-parallel to the
magnetic field, and because of the low excitation energy associated with
$m\rightarrow m+1$, more atoms can be added to H$_2$ to form larger
H$_N$ molecules. 

For a given magnetic field strength, as the number of H
atoms, $N$, increases, the electrons occupy more and more Landau orbitals
(with $m=0,~1,~2,\cdots,N-1$), and the transverse size of the molecule
increases as $R\sim (N/b)^{1/2}$. Let $a$ be the atomic spacing, and
$L_z\sim Na$ be the size of the molecule in the $z$-direction. 
The energy per atom in the molecule can be written as
$E\sim L_z^{-2}-l\,a^{-1}$, where $l=\ln (2a/R)$. 
Variation of $E$ with respect to $L_z$ gives 
\be
E\sim -N^2l^2,\qquad
L_z\sim Na \sim (Nl)^{-1}. 
\ee
This scaling behavior is valid for $1\ll N\ll N_s$.
The ``critical saturation number'' $N_s$ is reached when $a\sim R$, or
\be
N_s\sim b^{1/5},
\ee
(Lai \etal~1992). Beyond $N_s$, it becomes energetically more favorable
for the electrons to settle into the inner Landau orbitals (with smaller $m$)
with nodes in their longitudinal wavefunctions (i.e., $\nu\neq 0$). 
For $N\go N_s$, the energy per atom of the H$_N$ molecule,
$E=|E({\rm H}_N)|/N$, asymptotes to a value $\sim b^{2/5}$, 
and size of the atom scales as $R\sim a\sim b^{-2/5}$, independent of $N$
(see Sec.~V.A). For a typical magnetic field strength
($B_{12}=0.1-10^3$) of interest here, the energy saturation occurs
at $N_s\sim 2-6$ (see Fig.~4). 

\subsection{Intermediate Magnetic Field Regime}

For intermediate magnetic field strengths ($B\sim B_0\sim 10^9$~G), 
the only molecule that has been investigated in some detail is 
the hydrogen molecular ion H$_2^+$; both parallel (e.g., 
Le Guillou and Zinn-Justin 1984; Vincke and Baye 1985; 
Brigham and Wadehra 1987; Kappes and Schmelcher 
1995; Lopez \etal~1997;
Kravchenko and Liberman 1997) and nonparallel configurations (Larsen
1982; Wille 1987,1988; Kappes and Schmelcher 1996) have been studied using
different variational methods. 
Other one-electron molecular ions such as H$_3^{++}$ and H$_4^{+++}$
in strong magnetic fields have been studied by Lopez and Turbiner (2000)
(and references therein). For the H$_2$ molecule at intermediate fields,
earlier studies (e.g., Turbiner 1983; Basile \etal~1987) are of
qualitative character; quantitative calculations (for aligned
configurations) have been attempted only recently (Detmer \etal~1997,1998;
Kravchenko and Liberman 1998; Schmelcher \etal~2000). 

An important issue concerns the nature of the ground electronic state 
of H$_2$ as a function of $B$. Starting from the strong field regime
(see Sec.~IV.B and Sec.~IV.C) we know that for $B_{12}\go 0.2$ ($b\go 100$),
the ground state is the tightly bound state in which the electrons occupy
the $(m,\nu)=(0,0)$ and $(1,0)$ orbitals; this state corresponds to
$^3\Pi_u$ in the standard spectroscopic notation. For $b\lo 100$
the lowest energy state becomes the weakly bound state in which the electron
orbitals are $(0,0)$ and $(0,1)$; this corresponds to the $^3\Sigma_u$ state.
Detmer \etal~(1998) and Kravchenko and Liberman (1998)
found that for $b\lo 0.18$, the ground state is the usual $^1\Sigma_g$;
for $0.2\lo b\lo 12-14$, the ground state is $^3\Sigma_u$;
for $b\go 12$, the ground state is $^3\Pi_u$. 
However, their $^3\Sigma_u$ state is predominantly repulsive except for a very
shallow van-der-Waals (quadrupole-quadrupole interaction) minimum at large
proton separation. The binding of the $^3\Sigma_u$ was not demonstrated in
the calculations of Detmer \etal~(1998) and Kravchenko and Liberman (1998).
This is in contradiction to the $b\gg 1$ behavior of the
weakly bound state found in Lai and Salpeter (1996). Obviously, more work is
needed to attain a clear picture of how the energies of different H$_2$ states
behave as $B$ increases from $0$ to $b\gg 1$. 

\subsection{Molecules of Heavy Elements}

Molecules of heavy elements (other than hydrogen) in strong magnetic fields
have not been systematically investigated. 
There is some motivation to study 
molecules of light elements such as He, since if the hydrogen
on a neutron star surface is completely burnt out by nuclear reaction, 
helium would be the dominant species left in the atmosphere. There are also 
white dwarfs with pure He atmospheres. (Because of quick gravitational
separation of light and heavy elements in the gravitational field
of the neutron star or white dwarf, the atmosphere is expected to contain pure
elements.) For $b\gg 1$, the Hartree-Fock method discussed
in Sec.~IV.B can be generalized to the case of ion charge $Z>1$
(Lai \etal~1992).
Figure 5 shows the dissociation energy of the He$_2$ molecule as a function of
$B$. In general, we expect that, as long as $a_0/Z\gg (2Z-1)^{1/2}\hat\rho$,
or $b\gg 2Z^3$, the electronic properties of the heavy molecule is 
similar to those of H$_2$. When the condition $b\gg 2Z^3$ is not satisfied, 
the molecule should be quite different and may be unbound relative to 
individual atoms (e.g., Fe at $B=10^{12}$~G is unlikely to form a bound
molecule). Some Hartree-Fock results of diatomic molecules (from H$_2$
up to C$_2$) at $b=1000$ are given in Demeur \etal~(1994).

\section{Linear Chains and Condensed Matter}

As discussed in Sec.~IV.D, in a strong magnetic field
we can add more atoms to a diatomic molecule to form molecular chains.
When the number of atoms exceeds the saturation number, the structure of
the molecule is the same as that of an infinite chain. By placing a pile of
parallel chains together, three-dimensional condensed matter can be formed.

\subsection{Basic Scaling Relations for Linear Chains}

The simplest model for the linear chain is to treat it as a uniform cylinder
of electrons, with ions aligned along the magnetic field axis. 
After saturation, many electrons settle into the $\nu \neq 0$ states, and the
electrons can be treated as a Fermi sea in the $z$-direction. In order of
magnitude, the electrons occupy states with $m=0,1,2,\cdots,N_s-1$ and $\nu
=0,1,2,\cdots,N/N_s$. For $N\gg N_s \gg 1$, the uniform cylinder
approximation becomes increasingly valid because in the 
transverse direction the area covered by the $m$ Landau orbitals
scales as $(\sqrt {2m+1})^2 \propto m$; hence the volume 
increases with the number of interior electrons, giving a
constant electron density. Let $R$ be the radius of the cylinder and $a$
be the atomic spacing $a$ along the $z$-axis. The energy per atom 
(unit cell) in the chain can be written as (Ruderman 1971,1974; Chen 1973)
\be
E_\infty={2\pi^2Z^3\over 3b^2R^4a^2}-{Z^2\over a}
\left[\ln{2a\over R}-\left(\gamma-{5\over 8}\right)\right],
\label{einft}\ee
where $\gamma=0.5772\cdots$ is Euler's constant, and 
we have restored the dependence on the ion charge $Z$.
In Eq.~(\ref{einft}), the first term is the electron kinetic energy 
[see Eq.~(\ref{eqek})] and the second term is the (direct) Coulomb energy 
(the Madelung energy for the one-dimensional uniform lattice). 
Minimizing $E_\infty$ with respect to $R$ and $a$ gives
\ba
&&R=1.65\,Z^{1/5}b^{-2/5},\qquad a/R=2.14,\nonumber\\
&&E_\infty=-0.354\,Z^{9/5}b^{2/5}.
\label{chainscale}\ea
Not surprisingly, these scaling relations are the same as those for 
heavy atoms [see Eq.~(\ref{heavyatom})].

\subsection{Calculations of Linear Chains}

While the uniform cylinder model discussed above gives useful scaling relations
for the structure and energy of the linear chain, it is not sufficiently
accurate to determine the relative binding energy between the chain and the 
atom at $B_{12}\sim 0.1-100$. [The uniform cylinder model becomes 
accurate only when $N_s\sim b^{1/5}\gg 1$.] More refined calculations 
are needed to obtain accurate energies for field strengths characteristic 
of neutron star surfaces. Glasser and Kaplan (1975) generalized the uniform
cylinder model by considering quantized electron charge distribution
in the transverse direction. However, they assumed a uniform 
electron population in different Landau orbitals. The effect they treated 
amounts to only a small change in the value of the constant in the Madelung
energy expression, and therefore it is still insufficient to 
account for the binding of linear chains. The next step in a more relaxed 
variational calculation is to treat the effect of Coulomb potential 
on the population of electrons in different $m$ orbitals 
(Flowers \etal~1977; note that the calculation reported in this paper 
contained numerical errors, and was corrected by M\"uller 1984).
This is clearly an important ingredient for calculating the binding energy
of the chain since it allows for more electrons in the inner orbitals 
(small $m$'s), which increases the binding. A further improvement 
includes the nonuniform electron density distribution in the $z$-direction 
along the magnetic field (Neuhauser, Koonin and Langanke 1987). This effect 
is important for treating the bound electrons (i.e., the 
``electron core'' in Flowers \etal~1977) correctly for chains of heavy 
atoms like Fe. 

The self-consistent Hartree-Fock method for linear chains is similar to
that used for calculating multi-electron atoms
(see Sec.~III.C.2; Neuhauser \etal~1987). Consider a chain of length $Na$
(where $a$ is the ion spacing). The electron basis functions can be 
written as
\be
\Phi_{m\nu k}(\br)=W_m(\br_\perp){1\over\sqrt{N}}f_{m\nu}(z)\exp({ikz}),
\label{basis}\ee
where $k$ is the Bloch wavenumber, and $f_{m\nu}(z)=f_{m\nu}(z+a)$,
normalized via $\int_{-a/2}^{a/2}|f_{m\nu}(z)|^2dz=1$. In each
$(m\nu)$ band, the electrons occupy the $k$-space up to
$k_{m\nu}^F=\sigma_{m\nu}(\pi/a)$. Here
$\sigma_{m\nu}$ is the number of electrons in the $(m\nu)$ orbital
per unit cell in the chain, and satisfies the constraint
\be
\sum_{m\nu}\sigma_{m\nu}=Z.
\label{constraint}\ee
For each set of $\sigma_{m\nu}$, a coupled set of Hartree-Fock equations 
for $f_{m\nu}(z)$ can be derived, and the energy of the system determined. One
then varies $\sigma_{m\nu}$ and repeats the calculation until the energy
minimum is attained.

For linear chains consisting of light atoms such as H and He (or in general,
for sufficiently strong magnetic fields satisfying $b\gg 2Z^3$),
the electron density variation along $z$-axis is not significant since all the
electrons in the chain are ``ionized'' and are well approximated by plane
waves. Thus a variational calculation which assumes uniform density in the 
$z$-direction is adequate. This calculation is simpler than the full
Hartree-Fock calculation since the energy functional can be expressed in 
semi-analytic form (Lai \etal~1992). The basis electron wavefunctions are
given by Eq.~(\ref{basis}) with $f_{m\nu}=a^{-1/2}$. Electrons fill the
$m$-th orbital (band) up to a Fermi wavenumber given by $k_m^F=\sigma_m
(\pi/a)$, where $\sigma_m$ is the number of electrons in $m$-th orbital per
cell, with $m=0,1,2,\cdots,m_0-1$. The energy per cell in a chain can be
written as
\begin{eqnarray}
E = && {\hbar^2 \over 2 m_e} \sum_m \sigma_m 
{1 \over 3} \left(\sigma_m {\pi \over a}\right)^2 \nonumber \\
&&+ {Z^2 e^2 \over a}
\biggl [-\ln \left({2 a \over \hat \rho}\right) 
+ \gamma + {1 \over Z} \sum_m \sigma_m \psi (m+1)-
{1 \over 2 Z^2} \sum _{mm'} \sigma_m \sigma_{m'} Y_{mm'}\biggr ] \nonumber \\
&& -{e^2 \over 2 a} \sum_{mm'} \sigma_m \sigma_{m'}
\int_{-\infty}^{\infty}\! dz 
{\sin (\sigma_m \pi z \hat \rho /a) \over \sigma_m \pi z \hat \rho /a}
{\sin (\sigma_{m'} \pi z \hat \rho/a) \over \sigma_{m'} \pi z \hat \rho/a}
E_{mm'}(z),
\end{eqnarray}
where the three terms represent the kinetic energy, the direct 
Colulomb energy and the exchange energy. The digamma
function $\psi$ satisfies $\psi(m+1)=\psi(m)+(1/m)$, with $\psi(1)=-\gamma
=-0.5772\cdots$; the coefficient $Y_{mm'}$ depends on $m,~m'$;
the function $E_{mm'}(z)$ is the same as defined in Eq.~(\ref{eqemm}).
For a given lattice spacing $a$, the occupation numbers
$\sigma_m$ ($m=0,1,2,\cdots,m_0-1$) are varied to minimize the total 
energy $E$ under the constraint (\ref{constraint}) (with $\nu$ suppressed).
One can increase $m_0$ until further increase in $m_0$ results in no change 
in the distribution, i.e., $\sigma_{m_0-1}=0$. 
The constrained variation 
$\delta E - \eps_F \delta \sum_{m=0}^{m_0-1} \sigma _m =0$ yields
\begin{eqnarray}
\eps_F &=&{\hbar^2 \over 2 m_e} ({\pi \sigma_m \over a})^2 +
{Z^2 e^2 \over a} \biggl [{1 \over Z} \psi(m+1)-{1 \over Z^2}
\sum_{m'} \sigma_{m'} Y_{mm'} \biggr ] \nonumber \\
&& -{e^2 \over a} \sum_{m'} \int_{-\infty}^{\infty}\! dz
\cos (\sigma_m \pi z \hat \rho/a) 
{\sin (\sigma_{m'} \pi z \hat \rho/a) \over \sigma_{m'} \pi z \hat \rho/a} 
E_{mm'}(z). 
\label{efequal}\end{eqnarray}
Here $\eps_F$ is a constant Lagrange multiplier (Fermi energy) which 
must be determined self-consistently.
The system (\ref{efequal}) consists of $m_0$ equations for the $m_0$ unknown
parameters $\sigma_m$ plus the constant $\eps_F$.
They are solved together with Eq.~(\ref{constraint}) for these unknown 
quantities.

Density functional theory has also been used to calculate
the structure of linear chains in strong magnetic fields (Jones 1985; 
Relovsky and Ruder 1996). The problem with this approach is that the density
functional approximation has not been calibrated in strong magnetic fields, 
and therefore the accuracy of the approximation is not yet known (for a 
review of density functional theory as applied to non-magnetic terrestrial
solids, see, e.g., Callaway and March 1984). More accurate implementation of
the density functional theory in strong magnetic fields requires that the
current -- magnetic field interaction be taken into account (Vignale and 
Rasolt 1987,1988) and better exchange-correlation functional be used. 

\subsection{Cohesive Energy of Linear Chains}

Selected numerical results of the linear chains for a number of elements 
(up to $Z=26$) have been obtained by Neuhauser \etal~(1987) based on the
Hartree-Fock method and by Jones (1985) based on density functional 
theory, for a limited range of $B$'s around $10^{12}$~G. 
The numerical results for the energy 
(per atom) of the hydrogen chain (Lai \etal~1992) can be approximated by 
(to within $2\%$ accuracy for $1\lo B_{12} \lo 10^3$) 
\be
E_{\infty}({\rm H})
=-0.76\,b^{0.37}~({\rm a.u.})=-194\,B_{12}^{0.37}~({\rm eV}),
\label{hchain}\ee
This expression for $E_\infty$ is a factor of 1.8 times
that given in Eq.~(\ref{chainscale}). The cohesive energy of the 
H chain (energy release in H$+$H$_{\infty}=$H$_{\infty+1}$) is given (to
$\lo 10\%$ accuracy) by 
\be
Q_{\infty}^{(\infty)}({\rm H})=
|E_{\infty}({\rm H})|-|E(\rH)| \simeq 0.76\,b^{0.37}-0.16\,(\ln b)^2
\ee
where the superscript ``$(\infty)$'' indicates that the proton
has been treated as having infinite mass (see Sec.~V.E).
Figure 1 shows the cohesive energy of the H chain as a function
of $B$, and Table I gives some numerical values.
Figure 5 shows the similar result for the He chain. 
For these light elements, electron 
density variation along the $z$-axis can be safely ignored.

Numerical calculations carried out so far have indicated that 
for $B_{12}=1-10$, linear chains are unbound 
for large atomic numbers $Z\go 6$ (Jones 1986; Neuhauser \etal~1987).
In particular, the Fe chain is unbound relative to the Fe atom;
this is contrary to what some early calculations (e.g., Flowers 
\etal~1977) have indicated. 
Therefore, the chain-chain interaction must play a crucial role in
determining whether the three dimensional zero-pressure Fe condensed matter
is bound or not (see Sec.~V.E). The main difference between Fe and H 
is that for the Fe atom at $B_{12}\sim 1$, many electrons are populated 
in the $\nu\neq 1$ states, whereas for the H atom, as long as $b\gg 1$, 
the electron always settles down in the $\nu=0$ tightly bound state. Therefore,
the covalent bonding mechanism for forming molecules (see Sec.~IV.A) 
is not effective for Fe at $B_{12}\sim 1$. However, for a sufficiently 
large $B$, when $a_0/Z\gg \sqrt{2Z+1}\hat\rho$, or $B_{12}\gg
100(Z/26)^3$, we expect the Fe chain to be bound in a
similar fashion as the H chain or He chain.

\subsection{3D Condensed Matter: Uniform Electron Gas Model and its Extension}

A linear chain naturally attracts neighboring chains through
quadrupole-quadrupole interaction. By placing parallel chains close together
(with spacing of order $b^{-2/5}$), we obtain three-dimensional 
condensed matter (e.g., a body-centered tetragonal lattice; Ruderman 1971). 

The binding energy of the magnetized condensed matter at zero pressure can be 
estimated using the uniform electron gas model (e.g., Kadomtsev 1970).
Consider a Wigner-Seitz cell with radius $r_i=Z^{1/3}r_s$ ($r_s$ is the
mean electron spacing); the mean number density of electrons is $n_e=Z/(4\pi
r_i^3/3)$. The electron Fermi momentum $p_F$ is obtained from
$n_e=(eB/hc)(2p_F/h)$. When the Fermi energy $p_F^2/(2m_e)$ is less than the 
cyclotron energy $\hbar\omega_{ce}$, or when the electron number 
density satisfies
\be
n_e \le n_{B}
={1\over \sqrt{2}\pi^2\hat\rho^3}
=0.0716\,b^{3/2},
\label{nlandau}\ee
(or $r_i\ge r_{iB}=1.49\,Z^{1/3}b^{-1/2}$), the electrons only occupy
the ground Landau level. 
The energy per cell can be written as
\be
E_s(r_i)={3\pi^2Z^3\over 8b^2r_i^6}-{0.9Z^2\over r_i},
\label{estry}\ee
where the first term is the kinetic energy and the second term is the
Coulomb energy. For the zero-pressure condensed matter, we require
$dE_s/dr_i=0$, and the equilibrium $r_i$ and energy are then given by 
\ba
&&r_{i,0}\simeq 1.90\,Z^{1/5}b^{-2/5},\label{eqri0}\\
&&E_{s,0}\simeq -0.395\,Z^{9/5}b^{2/5}.\label{es0}
\ea
The corresponding zero-pressure condensation density is 
\be
\rho_{s,0}\simeq 561\,A\,Z^{-3/5}B_{12}^{6/5}\,{\rm g~cm}^{-3}
\label{rs0}\ee
Note that for $b\gg 1$, the zero-pressure density 
is much smaller than the ``magnetic'' density defined in 
Eq.~(\ref{nlandau}), i.e., $\rho_{s,0}/\rho_B=(r_B/r_{i,0})^3=
0.48\,Z^{2/5}b^{-3/10}$.

We now discuss several corrections to the uniform electron gas model.

(i) {\it Coulomb exchange interaction}.
The exclusion principle for the electrons results in an exchange
correction to the Coulomb energy. The Hartree-Fock exchange energy 
(in atomic units) per Wigner-Seitz cell is given by 
\be
E_{ex}=-{3Z\over 4\,b\,r_s^3}F,
\ee
where $F$ is a function of the ratio $y\equiv n_e/n_{B}$ 
(see Appendix A)
\be F=3-\gamma-2\ln (2y)-{2\over 3}\left[2\ln (2y)+\gamma 
-{13\over 6}\right] y^2+\cdots,
\ee
($\gamma=0.5772157\cdots$ is Euler's constant). 
The effect of this (negative) exchange interaction 
is to increase $r_{i,0}$ and $|E_{s,0}|$. 

(ii) {\it Relativistic effect}. As noted in Sec.~II,
the use of non-relativistic quantum mechanics for the bound states is a 
good approximation even for $B\go B_{\rm rel}\simeq 137^2 B_0$. We can show
that density-induced relativistic effect is also small.
The ``magnetic density'' $n_{B}$
for onset of Landau excitation is still given by Eq.~(\ref{nlandau}).
The relativistic parameter for the electron is 
$x_e\equiv p_F/(m_ec)=(n_e/n_{B})(2B/B_{\rm rel})^{1/2}$.
At the zero-pressure density as given by Eq.~(\ref{rs0}), we have 
$x_e\simeq 5\times 10^{-3} Z^{2/5}b^{1/5}$.
Thus near the zero-pressure density, relativistic effect is 
negligible for the range of magnetic field strengths of interest.

(iii) {\it Nonuniformity of the electron gas}. The Thomas-Fermi 
screening wavenumber $k_{\rm TF}$ is given by (e.g., Ashcroft and Mermin 1976)
$k_{\rm TF}^2=4\pi e^2 D(\eps_F)$, where $D(\eps_F)=
{\partial n_e/\partial \eps_F}$ 
is the density of states per unit volume at the Fermi surface 
$\eps=\eps_F=p_F^2/(2m_e)$. Since $n_e=(2eB/h^2c)p_F$, 
we have $D(\eps_F)={n_e/2\eps_F}=({m_e/n_e})({2eB/h^2c})^2$, and
\be
k_{\rm TF}=\left({4\over 3\pi^2}\right)^{1/2}b\,r_s^{3/2}.
\label{ktf}\ee
(More details on the electron screening in strong 
magnetic fields, including anisotropic effects, can be found in 
Horing 1969.) The gas is uniform when the screening length $k_{\rm TF}^{-1}$
is much longer than the particle spacing $r_i$, i.e., $k_{\rm TF}r_i\ll 1$. 
For the zero-pressure condensate with density parameter given by 
Eq.~(\ref{eqri0}), we have $k_{TF}r_i\simeq 1.83$, independent of $B$. Thus
even for $B\rightarrow \infty$, the nonuniformity of electron distribution
must be considered for the zero-pressure condensed matter. 
To leading order in $r_i\ll 1$, the energy correction (per cell) due to
nonuniformity can be calculated using linear response theory, which gives 
\be
E_{\rm TF}=
-{18\over 175}(k_{\rm TF}r_i)^2{(Ze)^2\over r_i}
\ee
(e.g., Lattimer \etal~1985; Fushiki \etal~1989). Using Eq.~(\ref{ktf}), we
have
\be
E_{\rm TF}=-0.0139\,Z\,b^2r_i^4.
\ee
Note that this expression is valid only for
$k_{\rm TF}r_i\ll 1$. At lower densities,
the nonuniformity effect can be studied only through detailed
electronic (band) structure calculations.
An approximate treatment relies on the Thomas-Fermi type statistical models, 
including the exchange-correlation and the Weizs\"acker gradient 
corrections (see Appendix B).

\subsection{Cohesive Energy of 3D Condensed Matter}

Although the simple uniform electron gas model and its Thomas-Fermi
type extensions give a reasonable estimate for the 
binding energy for the condensed state, they are not adequate for determining
the cohesive property of the condensed matter. The cohesive energy $Q_s$
is the difference between the atomic ground-state energy and the energy per
atom of the condensed matter ground state. 
One uncertainty concerns the lattice structure of
the condensed state, since the Madelung energy can be very different from
the Wigner-Seitz value [the second term in Eq.~(\ref{estry})]
for a non-cubic lattice. In principle, a three-dimensional
electronic band structure calculation is needed to
solve this problem, as Jones (1986) has attempted 
for a few elements using density functional theory. Jones adopted
a local approximation in one angular variable in solving the electron
eigenfunctions of the lattice Kohn-Sham potential; the validity of this 
approximation is not easy to justify. Moreover, 
without calibrations from other methods, the accuracy of density
functional approximation in a strong magnetic field is not known
(see discussion at the end of Sec.~V.B).

The energy difference $\Delta E_s=|E_{s,0}|-|E_\infty|$ between the
3d condensed matter and the 1d chain must be positive 
and can be estimated by calculating the interaction (mainly 
quadrupole--quadrupole) between the chains. Various considerations indicate
that the difference is between $0.4\%$ and $1\%$ of $|E_\infty|$
(Lai and Salpeter 1997). Therefore, for light elements such as
hydrogen and helium, the binding of the 3d condensed matter 
results mainly from the covalent bond along the magnetic field axis, 
not from the chain-chain interaction. For convenience we shall write
the cohesive energy $Q_s$ of the 3d hydrogen condensate in terms of the
cohesive energy ($Q_\infty$) of the linear chain as 
\be
Q_s(\rH)=Q_\infty(\rH)+\Delta E_s=(1+\zeta)Q_\infty(\rH),
\ee
with $\zeta\simeq 0.01-0.02$ for $B_{12}=1-500$. 

For hydrogen, the zero-point energy of the proton is not entirely
negligible, and can introduce a correction to the cohesive energy. 
The zero-point energy has not been rigorously calculated, but a reasonable
estimate is as follows. Neglecting the magnetic force,
the zero-point energy $E_{zp}$ of a proton in the the lattice is of order
$\hbar\Omega_p$, where $\Omega_p=(4\pi e^2n_e/m_p)^{1/2}$ 
is the proton plasma frequency. Using the the mean electron density 
$n_e\simeq 0.035\,b^{6/5}$ as estimated from Eq.~(\ref{chainscale}) or
Eq.~(\ref{eqri0}), we find
\be
E_{zp}\sim \hbar\Omega_p\simeq 0.015\,b^{3/5}.
\ee
This is much smaller than the total binding energy $|E_\infty|$ unless
$B_{12}\go 10^5$. This means that, for the range of field strengths of 
interest, the zero-point oscillation amplitude is small compared to the 
lattice spacing. Thus quantum melting is not effective (e.g., 
Ceperley and Alder 1980; Jones and Ceperley 1996),
and the condensed matter is a solid at zero temperature.
Accurate determination of $E_{zp}$ requires a detailed understanding 
of the lattice phonon spectra. At zero-field, Monte-Carlo simulations give 
$E_{zp}\simeq 3\hbar\Omega_p\eta/2$, with $\eta\simeq 0.5$ 
(Hansen and Pollock 1973). For definiteness, we will adopt the same
value for $E_{zp}$ in a strong magnetic field.
Taking into account the magnetic effect on the proton, the corrected
cohesive energy of the H chain is expected to be 
\be
Q_\infty= Q_\infty^{(\infty)}-{1\over 2}\left[\hbar\Omega_p\eta
+\hbar \left(\omega_{cp}^2+4\eta^2\Omega_p^2\right)^{1/2}-\hbar\omega_{cp}\right],
\label{eqqinfty}\ee
with $\eta\simeq 0.5$, where $\hbar\omega_{cp}=\hbar eB/(m_pc)$ 
is the cyclotron energy of the proton.  

Figure 1 depicts the cohesive energy of H$_\infty$ as a function of $B$;
the energy releases $Q_1$ and $Q_2$ for e+p=H and H+H=H$_2$ are also shown.
Some numerical values are given in Table I. The zero-point energy corrections
for $Q_2$ and $Q_\infty$ have been included in the figure (if they are
neglected, the curves are qualitatively similar, although 
the exact values of the energies are somewhat changed.)
Although $b\gg 1$ satisfies the nominal requirement for the
``strong field'' regime, a more realistic expansion
parameter for the stability of the condensed state over atoms
and molecules is the ratio $b^{0.4}/(\ln b)^2$. This ratio exceeds $0.3$, and 
increases rapidly with increasing field strength only for 
$b\go 10^4$. We see from Fig.~1 that $Q_1>Q_2>Q_\infty$ for $B_{12}\lo 10$,
and $Q_1>Q_\infty>Q_2$ for $10\lo B_{12}\lo 100$ and $Q_\infty>Q_1>Q_2$ 
for $B_{12}\go 100$. These inequalities have
important consequences on the composition of the saturated vapor
above the condensed phase for different magnetic fields (see Sec.~VII.B). 
Figure 5 shows similar numerical results for He.

The cohesive properties of the condensed state of heavy 
elements such as Fe are different from hydrogen or helium.
As discussed in Sec.~V.C, linear Fe chain is not bound relative to Fe
atom at $B_{12}=1-10$ (although we expect the Fe chain to be bound for
$B_{12}\gg 100$). Since chain-chain interactions only lower $E_s$
relative to $E_\infty$ by about $1\%$ (see above), it is 
likely that the 3d condensed state is also unbound. Jones (1996)
found a very small cohesive energy for the 3d condensed iron,
corresponding to about $0.5\%$ of the atomic binding energy. In view of 
the uncertainties associated with the calculations, Jones' results
should be considered as an upper limit, i.e.,
\be
Q_s\lo 0.005\,|E_{\rm atom}|\sim Z^{9/5}B_{12}^{2/5}~{\rm eV},\qquad
({\rm for}~~Z\go 10) 
\label{eqqs}\ee
where we have used Eq.~(\ref{eqeatom}) for $|E_{\rm atom}|$.

\subsection{Shape and Surface Energy of Condensed Droplets}

As we shall discuss in Sec.~VII, for sufficiently strong magnetic fields
and low temperatures, the condensed phase of hydrogen can be in pressure
equilibrium with the vapor phase. The two phases have markedly different 
densities and one might have an ``ocean/atmosphere interface''. The
question of droplets might have to be considered, and the shape and energy 
of a droplet is of interest (Lai and Salpeter 1997).  

For the phase equilibrium between the condensed state and the H$_N$
molecules in the vapor, the most relevant quantity is the 
``surface energy'' $S_N$, defined as the energy release in converting the 3d
condensate H$_{s,\infty}$ and a H$_N$ molecule into H$_{s,\infty+N}$. 
Clearly $S_1=Q_s=(1+\zeta)Q_\infty$ is the cohesive energy defined in 
Sec.~V.E. For a linear H$_N$ molecule with energy (per atom) $E_N=E({\rm
H}_N)$, we have 
\be
S_N=N (E_N-E_s)=N\Delta E_s+N(E_N-E_\infty)
=N\zeta Q_\infty+\xi Q_\infty,
\label{eqsn}\ee
where the first term on the right-hand side comes from cohesive
binding between chains, and the second term
is the ``end energy'' of the one-dimensional chain.
Based on numerical results for H$_2$,~H$_3$,~H$_4$ and H$_5$ (see Fig.~4),
we infer that the dimensionless factor $\xi$ in Eq.~(\ref{eqsn})
is of order unity. For $N\lo \xi/\zeta\sim 100$, the ``end energy''
dominates, while for $N\go 100$ the cohesion between chains becomes
important. In the latter case, the configuration which minimizes the
surface energy $S_N$ is not the linear chain, but
some highly elongated ``cylindrical droplet'' with $N_\perp$
parallel chains each containing $N_\parallel=N/N_\perp$ atoms.
For such a droplet, the ``end energy''
is of order $N_\perp \xi Q_\infty$. On the other hand there are
$\sim {N_\perp}^{1/2}$ ``unpaired'' chains in such a droplet,
each giving an energy $N_\parallel \zeta Q_\infty$. Thus the total
surface energy is of order $\left[N_\perp \xi +\zeta
(N/N_\perp){N_\perp}^{1/2}\right]Q_\infty$. The minimum surface energy
$S_N$ of the droplet, for a fixed $N\go 2\xi/\zeta\equiv N_c$, 
is then obtained for $N_\perp\simeq (N/N_c)^{2/3}$, $N_\parallel\simeq
N^{1/3}N_c^{2/3}$, and is of order 
\be
S_N\simeq 3\,\xi\left({N\over N_c}\right)^{2/3}\!\!Q_\infty,
\qquad {\rm for}~N\go N_c\equiv {2\xi\over\zeta}\sim 200.
\ee
Thus, although the optimal droplets are
highly elongated, the surface energy still
grows as $(N/200)^{2/3}$ for $N\go 200$. 


\section{Free Electron Gas in Strong Magnetic Fields}

In Sections III-V, we have reviewed the electronic structure and binding
energies of atoms, molecules and condensed matter in strong magnetic fields.
As discussed in Sec.~I.A, one of the main motivations to study matter 
in strong magnetic fields is to understand the neutron star surface 
layer, which directly mediates the thermal radiation 
from the star and acts as a boundary for the magnetosphere. 
Before discussing various properties of neutron star envelope in Sec.~VII, 
it is useful to summarize the basic thermodynamical properties
of a free electron gas in strong magnetic fields at finite temperature $T$.

The number density $n_e$ of electrons is related to the chemical potential
$\mu_e$ by 
\be
n_e={1\over (2\pi\hat\rho)^2\hbar}\sum_{n_L=0}^\infty g_{n_L}
\int_{-\infty}^\infty\! f\,\,dp_z,
\label{eqne}\ee
where $g_{n_L}$ is the spin degeneracy of the Landau level 
($g_0=1$ and $g_{n_L}=2$ for $n_L\ge 1$), and
$f$ is the Fermi-Dirac distribution
\be
f=\left[1+\exp\left({E-\mu_e\over kT}\right)\right]^{-1},
\ee
with $E$ given by Eq.~(\ref{eqrel}). The electron 
pressure is given by
\be
P_e={1\over (2\pi\hat\rho)^2\hbar}\sum_{n_L=0}^\infty g_{n_L}
\int_{-\infty}^\infty\! f\,\,{p_z^2c^2\over E}\,dp_z.
\label{eqpe}\ee
Note that the pressure is isotropic, contrary to what is stated
in Canuto and Ventura (1977) and some earlier papers
\footnote{The transverse kinetic pressure $P_{e\perp}$ 
is given by an expression similar to Eq.~(\ref{eqpe}), except that 
$p_z^2c^2$ is replaced by $\langle p_\perp^2c^2\rangle
= n_L\beta (m_ec^2)^2$. Thus the kinetic pressure is anisotropic,
with $P_{e\parallel}=P_e=P_{e\perp}+{\cal M}B$, where ${\cal M}$ is the
magnetization. When we compress the electron gas perpendicular to $\bB$
we must also do work against the Lorentz force density
$(\nabla\times {\bf\cal M})\times\bB$ involving the magnetization curent.
Thus there is a magnetic contribution to the perpendicular 
pressure of magnitude ${\cal M}B$. The composite pressure tensor is therefore
isotropic, in agreement with the thermodynamic result $P_e=-\Omega/V$ 
(Blandford and Hernquist 1982). For a nonuniform magnetic field,
the net force (per unit volume) on the stellar matter is
$-\nabla P_e-\nabla (B^2/8\pi)+(\bB\cdot\nabla)\bB/(4\pi)$.}. 
The grand thermodynamic potential is 
$\Omega=-P_e V$, from which all other
thermodynamic quantities can be obtained. Note that for 
nonrelativistic electrons (valid for $E_F\ll m_ec^2$ and $kT\ll m_ec^2$),
we use Eq.~(\ref{eqefree}) for $E$, and the expressions for 
the density $n_e$ and pressure $P_e$ can be simplified to
\ba
&&n_e={1\over 2\pi^{3/2}\hat\rho^2\lambda_{Te}}\sum_{n_L=0}^\infty
g_{n_L}I_{-1/2}\left({\mu_e-n_L\hbar\omega_{ce}\over kT}\right),\label{eqne2}\\
&&P_e={kT\over \pi^{3/2}\hat\rho^2\lambda_{Te}}\sum_{n_L=0}^\infty
g_{n_L}I_{1/2}\left({\mu_e-n_L\hbar\omega_{ce}\over kT}\right),
\ea
where $\lambda_{Te}\equiv (2\pi\hbar^2/m_ekT)^{1/2}$ is the thermal wavelength
of the electron, and $I_\eta$ is the Fermi integral:
\be
I_\eta(y)=\int_0^\infty\!\! {x^\eta\over \exp(x-y)+1}\,dx.
\ee

First consider degenerate electron gas at zero temperature. The Fermi energy
(excluding the electron rest mass) $E_F=\mu_e(T=0)-m_e c^2=(m_e c^2)
\epsilon_F$ is determined from
\be
n_e={\beta\over 2\pi^2\lambdabar_e^3}
\sum_{n_L=0}^{n_{\rm max}}g_{n_L}\,x_F(n_L),
\label{ef0}\ee
with
\be
x_F(n_L)={p_F(n_L)\over
m_ec}=\left[(1+\epsilon_F)^2-(1+2n_L\beta)\right]^{1/2},
\ee
where $\lambdabar_e=\hbar/(m_ec)$ is the electron Compton wavelength, 
$\beta=B/B_{\rm rel}=\alpha^2b$, and
$n_{\rm max}$ is set by the condition $(1+\epsilon_F)^2\ge (1+2n_{\rm max}
\beta)$. The electron pressure is given by
\be
P_e={\beta\, m_ec^2\over 2\pi^2\lambdabar_e^3}\sum_{n_L=0}^{n_{\rm max}}
g_{n_L}(1+2n_L\beta)\,\Theta\!\left[{x_F(n_L)\over (1+2n_L\beta)^{1/2}}\right],
\label{eqpe0}\ee
where
\be
\Theta(y)={1\over 2}y\sqrt{1+y^2}-{1\over 2}\ln\left(y+\sqrt{1+y^2}\right)
\longrightarrow {1\over 3}y^3~~{\rm for}~y\ll 1.
\ee
The critical ``magnetic density'' below which only the ground Landau
level is populated ($n_{\rm max}=0$) is determined by
$(1+\epsilon_F)^2=1+2\beta$, which gives [see Eq.~(\ref{nlandau})]
\be
\rho_B=0.802\,Y_e^{-1}
b^{3/2}~{\rm g~cm}^{-3}=7.04\times 10^3\,Y_e^{-1} B_{12}^{3/2}
~{\rm g~cm}^{-3},
\label{eqrhob}\ee
where $Y_e=Z/A$ is the number of electrons per baryon. Similarly, the critical
density below which only the $n_L=0,1$ levels are occupied ($n_{\rm max}=1$) 
is 
\be
\rho_{B1}=(2+\sqrt{2})\rho_B=3.414\,\rho_B.
\ee
For $\rho<\rho_B$, equation (\ref{ef0}) simplifies to 
\be
\rho=3.31\times 10^4\,Y_e^{-1} B_{12}\left[(1+\epsilon_F)^2-1\right]^{1/2}~{\rm
g~cm}^{-3}.
\ee
For nonrelativistic elelctrons ($\epsilon_F\ll 1$), the Fermi temperature
$T_F=E_F/k=(m_e c^2/k)\epsilon_F$ is given by
\be
T_F={E_F\over k}= 2.70\, B_{12}^{-2}\left({Y_e\,\rho}\right)^2~{\rm K}
\qquad ({\rm for}~~\rho<\rho_B),
\label{eqtf}\ee
where $\rho$ is in units of 1~g~cm$^{-3}$.
For $\rho\gg\rho_B$, many Landau levels are filled by the electrons, 
Eqs.~(\ref{eqne}) and (\ref{eqpe}) reduce to the zero-field expressions. 
In this limit, the Fermi momentum $p_F$ is given by
\be 
x_F={p_F\over m_ec}={\hbar\over m_ec}(3\pi^2n_e)^{1/3}=
1.009\times 10^{-2}\left({Y_e\,\rho}\right)^{1/3},\qquad (B=0)
\label{eqxf}\ee
and the Fermi temperature is 
\be
T_F={m_ec^2\over k}\left(\sqrt{1+x_F^2}-1\right)\simeq 3.0\times 10^5
\left({Y_e\,\rho}\right)^{2/3}~{\rm K},\qquad (B=0)
\label{eqtf0}\ee
where the second equality applies to nonrelativistic electrons ($x_F\ll 1$). 
Comparison between Eq.~(\ref{eqtf}) and Eq.~(\ref{eqtf0}) 
clearly shows that the magnetic field lifts the degeneracy of
electrons even at relatively high density (see Fig.~6).

Finite temperature tends to smear out Landau levels. 
Let the energy difference between the $n_L=n_{\rm max}$ level and the
$n_L=n_{\rm max}+1$ level be $\Delta E_B$. We can define a ``magnetic
temperature'' 
\be
T_B={\Delta E_B\over k}={m_e c^2\over k}\left(\sqrt{1+2n_{\rm max}\beta
+2\beta}-\sqrt{1+2n_{\rm max}\beta}\right).
\label{eqtb}\ee
Clearly, $T_F=T_B$ at $\rho=\rho_B$ (see Fig.~6). 
The effects due to Landau quantization are diminished when $T\go T_B$. 
For $\rho\le \rho_B$, we have
$T_B=\left(\sqrt{1+2\beta}-1\right)(m_ec^2/k)$, which reduces to
$T_B\simeq \hbar\omega_{ce}/k$ for $\beta=\alpha^2b \ll 1$. 
For $\rho\gg \rho_B$ (or $n_{\rm max}\gg 1$), equation (\ref{eqtb})
becomes
\be
T_B\simeq {\hbar\omega_{ce}\over k}\left({m_e\over m_e^\ast}\right) 
=1.34\times 10^8\,B_{12}\,(1+x_F^2)^{-1/2}~{\rm K},
\ee
where $m_e^\ast=\sqrt{m_e^2+(p_F/c)^2}=m_e\sqrt{1+x_F^2}$, with
$x_F$ given by Eq.~(\ref{eqxf}). 

There are three regimes characterizing the effects of Landau quantization 
on the thermodynamic properties of the electron gas (Figure 6; see also 
Yakovlev and Kaminker 1994):

(i) $\rho\lo\rho_B$ and $T\lo T_B$: In this regime, the electrons 
populate mostly the ground Landau level, and the magnetic field modifies 
essentially all the properties of the gas. The field is sometimes 
termed ``strongly quantizing''. For example, for degenerate, nonrelativistic
electrons ($\rho<\rho_B$ and $T\ll T_F\ll m_ec^2/k$),
the internal energy density and pressure are
\ba
&&u_e={1\over 3}n_eE_F,\\
&&P_e=2u_e={2\over 3}n_eE_F\propto B^{-2}\rho^3.
\label{eqpe1}\ea
These should be compared with the $B=0$ expression 
$P_e=2u_e/3\propto\rho^{5/3}$.
Note that for nondegenerate electrons ($T\gg T_F$), the classical ideal gas
equation of state, 
\be
P_e=n_ekT,
\ee
still holds in this ``strongly quantizing''
regime, although other thermodynamic quantities are 
significantly modified by the magnetic field.

(ii) $\rho\go\rho_B$ and $T\lo T_B$: In this regime,
the electrons are degenerate (note that $T_F>T_B$ when $\rho>\rho_B$; see
Fig.~6), and populate many Landau levels but the level spacing exceeds
$kT$. The magnetic field is termed ``weakly quantizing''. 
The bulk properties of the gas (e.g., pressure and chemical potential),
which are determined by all the electrons in the Fermi sea, are only slightly
affected by such magnetic fields. However, 
the quantities determined by thermal electrons 
near the Fermi surface show large oscillatory features as a function of
density or magnetic field strength. These de Haas - van Alphen type
oscillations arise as successive Landau levels are occupied with 
increasing density (or decreasing magnetic field).
The oscillatory quantities are usually expressed as derivatives of the 
bulk quantities with respect to thermodynamic variables; 
examples include heat capacity, magnetization and magnetic susceptibility,
adiabatic index $(\partial\ln P_e/\partial\ln\rho)$, sound speed, 
and electron screening length of an electric charge in the plasma 
(e.g., Ashcroft and Mermin 1976; Blandford and Hernquist 1982; 
Lai and Shapiro 1991; Yakovlev and Kaminker 1994).
With inceasing $T$, the oscillations become weaker because of
the thermal broadening of the Landau levels; when $T\go T_B$, the oscillations
are entirely smeared out, and the field-free results are recovered.
 
(iii) For $T\gg T_B$ (regardless of density): In this regime, many Landau
levels are populated and the thermal widths of the Landau levels ($\sim kT$)
are higher than the level spacing. The magnetic field is termed 
``non-quantizing'' and does not affect the thermodynamic properties of 
the gas.

\section{Surface Layer of a Magnetized Neutron Star}

In this section we review the properties of the 
surface layer of a magnetized neutron star (NS).
We shall focus on the thermodynamic property and phase diagram.  
We expect various forms of magnetic bound states discussed in Sections III-V
to exist on the stellar envelope depending on the field strength, 
temperature and density.

The chemical composition of the NS surface is unknown. 
A NS is formed as a collapsed, hot ($kT\go 10$~MeV) core
of a massive star in a supernova explosion. The NS matter
may be assumed to be fully catalyzed and in the lowest energy 
state (e.g., Salpeter 1961; Baym, Pethick and Sutherland 1971).  
We therefore expect the NS surface to consist
of iron ($^{56}$Fe) formed at the star's birth.
This may be the case for young radio pulsars that have not accreted any gas.
However, once the NS accretes material, or has gone through a phase
of accretion, either from the interstellar medium or from a binary companion, 
the surface (crust) composition can be quite different due to surface 
nuclear reactions and weak interactions during the accretion (e.g., Haensel and
Zdunik 1990; Blaes \etal~1990; Schatz \etal~1999). Moreover,
a hydrogen-helium envelope will form on the top of the surface unless it has 
completely burnt out. While a strong magnetic field and/or rapid stellar 
spin may prevent large-scale accretion, 
it should be noted that even with a low accretion rate 
of $10^{10}$ g~s$^{-1}$ (typical of accretion from the interstellar
medium) for one year, the accreted material will be more 
than enough to completely shield the original iron surface of the NS.
The lightest elements, H and He, are likely to be the 
most important chemical species in the envelope due to their 
predominance in the accreting gas and also due to quick 
separation of light and heavy elements in the gravitational field 
of the NS (see Alcock and Illarionov 1980 for a discussion 
of gravitational separation in white dwarfs; applied to NSs, we find
that the settling time of C in a $10^6$ K hydrogen 
photosphere is of order a second). If the present accretion rate is low, 
gravitational settling produces a pure H envelope. Alternatively, a
pure He layer may result if the hydrogen has completely burnt out. 

In this section, we shall mostly focus on the hydorgen envelope 
(Sec.~VII.A and Sec.~VII.B) both because of its predominance in the 
outer layer of the NS and because the properties of different phases of 
H are better understood. A pure He envelope would presumably have similiar
properties as the H envelope. The Iron surface layer and deeper crust will be 
discussed in Sec.~VII.C and Sec.~VII.E. We are only interested in the 
temperature regime $T\go 10^5$~K, since NSs with 
$T\lo 10^5$~K are nearly impossible to observe.

\subsection{Warm Hydrogen Atmosphere}

We now consider the physical conditions and chemical equilibrium 
in a hydrogen atmosphere with photospheric temperature in the range
$T_{\rm ph}\sim 10^5-10^{6.5}$~K and magnetic field strength in the range 
$B_{12}\sim 0.1-20$. These conditions are likely to be satisfied by 
most observable NSs. For such relatively low field strengths, 
the atmosphere is largely nondegenerate, and consists mainly of 
ionized hydrogen, H atoms and small H$_N$ molecules, and we can neglect the
condensed phase in the photosphere. 
(In Sec.~VII.B we shall consider the more extreme situation of 
$B_{12}\gg 10$ in which the nondegenerate atmosphere has negligible
optical depth and the condensed phase becomes important.)
Although the density scale height of the
atmosphere is only $h\simeq k T_{\rm ph}/ m_p g \simeq 0.08\,T_{\rm
ph,5}\,g_{14}^{-1}$~cm, where $T_{\rm ph,5}=T_{\rm ph}/(10^5~{\rm K})$, and
$g=10^{14}\,g_{14}$~cm~s$^{-2}$ is the gravitational acceleration,
the atmosphere has significant optical depth and therefore the atmospheric
properties determine the thermal radiation spectrum from the NS
(see Pavlov \etal~1995 and references therein).

The photosphere of the NS is located at the characteristic photon optical 
depth $\tau=\int_{r_{\rm ph}}^\infty\rho\,\kappa_R\,dr=2/3$, where $\kappa_R$
(in units of cm$^2/$g) is the Rosseland mean opacity;
the photosphere pressure is $P_{\rm ph}\simeq 2g/(3\kappa_R)$.
An accurate determination of the photosphere conditions requires
a self-consistent solution of the atmosphere structure and radiative
transport, but an order-of-magnitude estimate is as follows.  
In a strong magnetic field, the radiative opacity becomes anisotropic
and depends on polarization (e.g., Canuto \etal~1971; Lodenquai \etal~1974; 
Pavlov and Panov 1976; Ventura 1979; Nagel and Ventura 1980; see Meszaros 1992 
and references therein).
For photons with polarization vector perpendicular to the
magnetic field (the ``extraordinary mode'') the free-free absorption and
electron scattering opacities are reduced below their zero-field values by a
factor $(\omega/\omega_{ce})^2$, while for photons polarized along the magnetic
field (the ``ordinary mode''), the opacities are not affected. Pavlov and
Yakovlev (1977) and Silant'ev and Yakovlev (1980) have calculated the
appropriate averaged Rosseland mean free-free and scattering opacity; 
in the magnetic field and temperature 
regime of interest, an approximate fitting formula is 
\be
\kappa_R(B)\simeq  400\,\eta\left({kT\over\hbar\omega_{ce}}\right)^2\kappa_R
(0),
\ee
where $\eta\simeq 1$ and $\kappa_R(0)$ is the zero-field opacity.
Using the ideal gas equation of state, $P_{\rm ph}\simeq \rho_{\rm ph}
kT_{\rm ph}/m_p$, we obtain the photosphere density
\be
\rho_{\rm ph} \simeq 0.5\,\eta^{-1/2}\,g_{14}^{1/2}
\,T_{\rm ph,5}^{1/4}\,B_{12}~{\rm g\,cm}^{-3}.
\ee
Note that for $\eta\sim 1/400$, this equation
also approximately characterizes the density of the
deeper layer where the extraordinary photons are emitted. 
Other sources of opacity such as bound-free and bound-bound absorptions
will increase the opacity and reduce the
photosphere density, but the above estimate defines the general 
range of densities in the atmosphere if the $T_{\rm ph}$ is large 
enough for the neutral H abundance to be small. 
Clearly, a typical atmosphere satisfies $\rho\ll \rho_B$ and
$T\ll T_B$, but $T\gg T_F$ (see Sec.~VI and Fig.~6), i.e., 
the magnetic field is strongly quantizing in the atomsphere, but
the electrons are nondegenerate. 

An important issue for NS atmosphere modeling concerns
the ionization equilibrium (or Saha equilibrium)
of atoms in strong magnetic fields.
Earlier treatments of this problem (e.g., Gnedin \etal~1974;
Khersonskii 1987) assumed that the atom can move freely 
across the magnetic field; this is generally not valid for the 
strong field regime of interest here (see Sec.~III.E). 
Lai and Salpeter (1995) gave an approximate analytic solution
for a limited temperature-denisty regime (see below).
To date the most complete treatment of the problem is that of Potekhin et
al.~(1999), who used the numerical energy levels of a moving H atom as
obtained by Potekhin (1994,1998) and an approximate description of the nonideal
gas effect to derive the thermodynamic properties of a 
partially ionized hydrogen
plasma in strong magnetic fields. Here we discuss the basic issues of
ionization equilibrium in strong magnetic fields, and refer to
Potekhin \etal~(1999) for a more detailed treatment.

For nondegenerate electrons in a magnetic field, the partition function
(in volume $V$) is 
\ba
Z_e&=&{V\over 2\pi\hat\rho^2}\sum_{n_L=0}^\infty g_{n_L} 
\exp\left({-n_L\hbar\omega_{ce}
\over kT}\right)\int_{-\infty}^\infty{dp_z\over h}
\exp\left({-p_z^2\over 2m_ekT}\right)\nonumber\\
&=&{V\over 2\pi\hat\rho^2\lambda_{Te}}\tanh^{-1}\left({\hbar\omega_{ce}
\over 2kT}\right)\nonumber\\
&\simeq& {V\over 2\pi\hat\rho^2\lambda_{Te}},
\label{eqze}\ea
where $\lambda_{Te}=(2\pi\hbar^2/m_ekT)^{1/2}$ is the electron
thermal wavelength, and the last equality applies for $T\ll T_B$.
For protons, we shall drop the zero-point energy and the spin energy
in both free states and bound states [i.e., $E=n_L\hbar\omega_{cp}
+p_z^2/(2m_p)$ for a free proton]. Treating the proton as a spinless
particle ($g_{n_L}=1$), we find that the partition function of free protons is
\be
Z_p={V\over 2\pi\hat\rho^2\lambda_{Tp}}\left[1-\exp\left(-{\hbar\omega_{cp}
\over kT}\right)\right]^{-1},
\ee
where $\lambda_{Tp}=(2\pi\hbar^2/m_pkT)^{1/2}$ is the proton thermal
wavelength. 

Using Eq.~(\ref{eqemtot}) for the energy of the H atom, we write the 
partition function for the bound states as
\be
Z_H={V\over h^3}\sum_{m\nu}\int\!\!d^3K\,w_{m\nu}(K_\perp)
\exp\left(-{{\cal E}_{m\nu}\over kT}
\right)={V\over\lambda_{TH}^3}Z_w,
\label{eqzh}\ee
where $\lambda_{TH}=(2\pi\hbar^2/MkT)^{1/2}$, and
\ba
&&Z_{m\nu}={\lambda_{TH}^2\over 2\pi\hbar^2}
\int\!\! dK_\perp\,K_\perp w_{m\nu}(K_\perp)\exp\left({-{m\hbar\omega_{cp}
+E_{m\nu}(K_\perp)\over kT}}\right),\label{eqzmn}\\
&&Z_w=\sum_{m\nu}Z_{m\nu}.
\ea
The Saha equation for the ionization equilibrium in strong
magnetic fields then reads
\be
{n_H\over n_p n_e}={V Z_H\over Z_eZ_p}=
{\lambda_{Te}\lambda_{Tp}(2\pi\hat\rho^2)^2\over \lambda_{TH}^3}
\left[1-\exp\left(-{\hbar\omega_{cp}\over kT}\right)\right]Z_w.
\label{eqsaha}\ee

In Eqs.~(\ref{eqzh})-(\ref{eqzmn}), $w_{m\nu}(K_\perp)$ is the 
occupation probability of the hydrogen bound state characterized by
$m,\nu,K_\perp$, and it measures deviation from Maxwell-Boltzmann distribution
due to the medium effect. Physically, it arises from the fact that an atom
tends to be ``destroyed'' when another particle in the medium comes close
to it. The atom--proton interaction introduces
a correction to the chemical potential of the atomic gas
\be
\Delta\mu_H\simeq 2n_pkT\!\int\!\!d^3r\,\left[1-\exp(-U_{12}/kT)\right],
\ee
(see, e.g., Landau and Lifshitz 1980), where $U_{12}$ is the interaction 
potential. Similar expressions can be written down for atom--electron
and atom--atom interactions. For $r$ much larger than the size
of the elongated atom, the atom--proton potential $U_{12}$ has the form
$U_{12}\sim eQ(3\cos^2\theta-1)/r^3$, where $\theta$ is the angle between
the vector ${\bf r}$ and the $z$-axis, and $Q\sim eL_z^2$ is the quadrupole 
moment of the atom; the atom--atom interaction potential has the form
$U_{12}\sim Q^2(3-30\cos^2\theta+35\cos^4\theta)/r^5$.
Since the integration over the solid angle $\int\!d\Omega\, U_{12}=0$ at large
$r$, the contribution to $\Delta\mu(\rH)$ from large $r$
is negligible. (An atom with $K_\perp\ne 0$ also acquires a dipole
moment in the direction of $\bB\times \bK_\perp$, the resulting dipole
interaction also satisfies $\int\!d^3r\,U_{12}=0$.)
Thus the main effect of particle interactions is the ``excluded volume
effect'': Let $L_{m\nu}(K_\perp)$ be the characteristic size of the atom
such that we can set $U_{12}\rightarrow\infty$ when $r\lo L_{m\nu}$. 
We then have $\Delta\mu_H\sim n_bkT (4\pi L_{m\nu}^3/3)$, where $n_b
=n_H+n_p$ is the baryon number density. Therefore, 
the occupation probability is of order
\be
w_{m\nu}(K_\perp)\sim \exp\left[-{4\pi\over 3}n_bL_{m\nu}^3(K_\perp)\right].
\ee
The size of the atom can be estimated as follows
(see Sec.~III.A and Sec.~III.E):  when $K_\perp=0$, the tightly bound state
($\nu=0$) has $L_z\sim l_m^{-1},~L_\perp\sim\rho_m$, while the $\nu>0$ state
has $L_z\sim \nu^2,~L_\perp\sim\rho_m$; when $K_\perp\ne 0$, 
the electron and proton are displaced in the transverse direction by 
a distance $d\lo K_\perp/b$; thus we have $L_{m\nu}(K_\perp)\sim
\max(L_z,L_\perp,K/b)$. More accurate fitting formula for $L_{m\nu}(K_\perp)$
is given in Potekhin (1998). 

More precise calculation of the occupation probability requires 
detailed treatment of interactions between various particles in the 
plasma. Even at zero magnetic field, the problem is challenging
and uncertainties remain (e.g., Hummer and Mihalas 1988; Mihalas et
al.~1988; Saumon and Chabrier 1991,1992; Potekhin 1996). In strong magnetic
fields, additional complications arise from the nonspherical shape of 
the atom. Nevertheless, using the hard-sphere approximation similar to
the field-free situation, Potekhin \etal~(1999) have constructed 
a free energy model for partially ionized hydrogen, from which
$w_{m\nu}(K_\perp)$ can be calculated along with other thermodynamic 
quantities. Figure 7 shows some numerical results based on the
calculations of Potekhin \etal~(1999).
The general trend that the neutral fraction $n_H/n_b$ increases with increasing
$B$ is the result of larger binding energy $|E_{m\nu}(K_\perp)|$ for larger 
$B$, although this trend is offset partially by the fact that
the electron phase space also increases with $B$ [see Eq.~(\ref{eqze})].
For a given $T$ and $B$, the neutral fraction increases with density
until $\rho$ reaches $\rho_c\sim 10$-100~g~cm$^{-3}$ [for $B=10^{12}$
-$10^{13}$~G (see Potekhin \etal~1999); 
$\rho_c$ scales with $b$ roughly as $(\ln b)^3$], above which the neutral
fraction declines because of pressure ionization.

It is instructive to consider the relative importance of the ``centered
states'' and the ``decentered states'' to the neutral hydrogen fraction. 
As discussed in Sec.~III.E, the H energy can be approximated by
$E_0(K_\perp)\simeq E_0+K_\perp^2/(2M_{\perp})$ for
$K_\perp\lo K_{\perp c}\sim \sqrt{2M|E_0|}$ (we consider 
only the $m=\nu=0$ state for simplicity). Setting $w_{00}(K_\perp)=1$
for low densities, the contribution of the $K_\perp\lo K_{\perp c}$ states 
to $Z_w$ is given by (in a.u.)
\be 
Z_w^{(c)}\simeq \exp\left(-{E_{0}\over
T}\right)\left({M_{\perp}\over M}\right),
\ee
for $K_{\perp c}^2/(2M_\perp)\sim (M/M_\perp)|E_{0}|\gg T$.
On the other hand, the contribution from the decentered states 
to $Z_w$ can be estimated as
\be
Z_w^{(d)}\simeq {1\over MT}\int_{K_\perp c}^{K_{\rm max}}
\! K_\perp\exp\left[-{E_{0}(K_\perp)\over T}\right] dK_\perp,
\ee
where $K_{\rm max}$ is given by $(4\pi/3)(K_{\rm max}/b)^3n_b\simeq 1$.
Since $|E_0(K_\perp)|$ for $K\go K_{\perp c}$ is much less than $|E_0|$,
we find 
\be
{Z_w^{(c)}\over Z_w^{(d)}}\sim
{2M_\perp T\over K_{\rm max}^2}\exp\left({|E_0|\over T}\right)
\sim {5M_\perp T\over b^2}n_b^{2/3}\exp\left
({|E_0|\over T}\right).
\ee
Thus for the centered states to dominate the atomic population,
we require $|E_0|/T\go \ln\left(b^2/5M_\perp Tn_b^{2/3}\right)$;
for example, at $T=10^{5.5}$~K,  this corresponds to $B\go 10^{13.5}$~G,
or $|E_0|/T\go 15$ (see the dotted lines in Fig.~7).

For sufficiently high $B$ or sufficiently low $T$, we expect the 
atmosphere to have appreciable abundance of H$_2$ molecules. 
Since detailed study of the effect of motion on H$_2$ is not available,
only an approximate estimate is possible (Lai and Salpeter 1997; 
Potekhin \etal~1999). For example, at $B=10^{13}$~G there exists
a large amount of H$_2$ in the photosphere ($\rho\sim 10-100$~g~cm$^{-3}$)
when $T\lo 10^{5.5}$~K. As we go deeper into the atmosphere,
we expect larger molecules to appear; when the density approaches
the internal density of the atom/molecule, the bound states lose
their identities and we obtain a uniform, ionized plasma.
With increasing density, the plasma becomes degenerate, and
gradually transforms into a condensed phase
(Lai and Salpeter 1997).

\subsection{Surface Hydrogen at Ultrahigh Fields: The Condensed Phase}

We have seen in Sec.~VII.A that for $T\go 10^5$~K and $B\lo 10^{13.5}$~K, 
the outermost layer of the neutron star is nondegenerate, and the surface
material gradually transforms into a degenerate Coulomb plasma as density
increases. As discussed in Sections III-V, the binding energy of the condensed
hydrogen increases as a power-law function of $B$, while the binding energies
of atoms and small molecules increase only logarithmically. 
We therefore expect that for sufficiently strong magnetic fields, there exists
a critical temperatire $T_{\rm crit}$, below which a first-order phase 
transition occurs between the condensed hydrogen and the gaseous vapor;
as the vapor density decreases with increasing $B$ or decreasing $T$,
the outermost stellar surface would be in the form of condensed hydrogen
(Lai and Salpeter 1997). 

While a precise calculation of $T_{\rm crit}$ for the phase 
transition is not available at present, we can get an estimate by 
considering the equilibrium between the condensed hydrogen (labeled ``s'')
and the gaseous phase (labeled ``g'') in the ultrahigh field regime (where
phase separation exists). The gaseous phase consists of a mixture of
free electrons, protons, bound atoms and molecules. Phase equilibrium 
requires the temperature, pressure and the
chemical potentials of different species to satisfy the conditions
\ba
&&P_s=P_g=[2n_p+n(\rH)+n(\rH_2)+n(\rH_3)+\cdots]kT=P,\\
&&\mu_s=\mu_e+\mu_p=\mu(\rH)={1\over 2}\mu(\rH_2)={1\over 3}\mu(\rH_3)
=\cdots
\ea
For the condensed phase near zero-pressure, the density is approximately
\be
\rho_s\simeq 561\,B_{12}^{6/5}~{\rm g~cm}^{-3},
\ee
and the electron Fermi temperature
is $T_F\simeq 0.236\,b^{2/5}=8.4\times 10^5B_{12}^{2/5}$~K; 
thus at a given temperature, the condensed hydrogen becomes 
more degenerate as $B$ increases. Let the energy per Wigner-Seitz
cell in the condensate be $E_s(r_s)$ [see Eq.~(\ref{estry}) for an approximate
expression; $r_s=r_i$ for hydrogen]. The pressure and chemical potential of the
condensed phase are given by 
\ba
&&P_s=-{1\over 4\pi r_s^2}{dE_s\over dr_s},\\
&&\mu_s=E_s(r_s)+P_sV_s\simeq E_{s,0}+P_sV_{s,0},
\ea
where the subscript ``$0$'' indicates the zero-pressure values. We have assumed
that the vapor pressure is sufficiently small so that the deviation from the
zero-pressure state of the condensate is small, i.e., $\delta\equiv
|(r_s-r_{s,0})/r_{s,0}|\ll 1$; this is justified when the saturation vapor
pressure $P_{\rm sat}$ is much less than the critical pressure $P_{\rm crit}$
for phase separation, or when $T\ll T_{\rm crit}$. The finite temperature
correction $\Delta\mu_s$ to the chemical potential of the condensed phase,
as given by $\Delta\mu_s(T)\simeq \pi^2T^2/(12T_F)$, is much smaller than
the cohesive energy and can be neglected.
Using the partition functions of free electrons, protons and atoms
as given in Sec.~VII.A, we find that in the saturated vapor,
\ba
&&n_p=n_e\simeq {bM^{1/4}T^{1/2}\over (2\pi)^{3/2}}
\left[1-\exp\left(-{b\over MT}\right)\right]^{-1/2}\!
\exp\left(-{Q_1+Q_s\over 2T}\right),\label{eqnp}\\
&&n(\rH)\simeq \left({MT\over 2\pi}\right)^{3/2}
\!\left({M_\perp'\over M}\right)\exp\left(-{Q_s\over T}\right).
\label{eqnh}
\ea
where $Q_s=|E_{s,0}|-|E(\rH)|$ is the cohesive energy of the condensed 
hydrogen (we have neglected $P_sV_{s,0}$ in comparison to $Q_s$),
and $Q_1=|E(\rH)|$ is the ionization energy of the hydrogen atom.
In Eq.~(\ref{eqnh}) we have included only the ground state ($m=\nu=0$)
of the H atom and have neglected the ``decentered states''; this is valid
for $T\lo Q_1/20$ (see Sec.~VII.A). The equilibrium condition 
$N\mu_s=\mu_N$ for the process $\rH_{s,\infty}+H_N=\rH_{s,\infty+N}$ 
(where H$_N$ represents a small molecular chain or a 3d droplet) yields
\be
n(\rH_N)\simeq N^{3/2}\left({MT\over2\pi}\right)^{3/2}
\exp\left(-{S_N\over T}\right),
\label{eqhn}\ee
where $S_N=NE_N-NE_s$ is the surface energy discussed in Sec.~V.F.
In Eq.~(\ref{eqhn}), we have assumed that the H$_N$ molecule (or 3d droplet) 
moves across the field freely; this should be an increasingly good 
approximation as $N$ increases.

The critical temperature $T_{\rm crit}$, below which phase 
separation between the condensed hydrogen and the gaseous vapor occurs, is
determined by the condition $n_s=n_g=n_p+n(\rH)+2n(\rH_2)+3n(\rH_3)+\cdots$. 
Although Eqs.~(\ref{eqnp})-(\ref{eqhn}) are derived for $n_g\ll n_s$,
we may still use them to obtain an estimate of $T_{\rm crit}$.
Using the approximate surface energy $S_N$ as discussed in Sec.~V.F,
we find
\be
T_{\rm crit}\sim 0.1\,Q_s\simeq 0.1\,Q_\infty.
\ee
Thus $T_{\rm crit}\simeq 8\times 10^4$, $5\times 10^5$ and $10^6$~K for
$B_{12}=10,~100$ and $500$. Figure 8 shows some examples of
the saturation vapor density as a function of temperature for several
values of $B$. It should be emphasized that the calculation is very uncertain
around $T\sim T_{\rm crit}$. But when the temperature is below
$T_{\rm crit}/2$ (for example), the valor density becomes much less than
the condensation density $n_s$ and phase transition is unavoidable.
When the temperature drops below a fraction of $T_{\rm crit}$,
the vapor denisty becomes so low that the optical depth of the vapor is
negligible and the outermost layer of the NS then consists of
condensed hydrogen. The condensate will be in the liquid state when
$\Gamma=e^2/(r_ikT)\lo 175$ (see Sec.~VII.E), or when
$T\go 1.3\times 10^3\,(\rho/1~{\rm g~cm}^{-3})^{1/3}\,{\rm K}
\simeq 7\times 10^4\,B_{14}^{2/5}$~K. The radiative properties of such
condensed phase are of interest to study (see Sec.~VII.D).
 

The protons in the condensed hydrogen phase can undergo significant
pycnonuclear reactions. Unlike the usual situation of 
pycronuclear reactions (see Salpeter and Van Horn 1969; Ichimaru 1993),
where the high densities needed for the reactions at low temperatures
are achieved through very high pressures, here the large densities
(even at zero pressure) result from the strong magnetic field 
--- this is truely a ``zero-pressure cold fusion'' (Lai and Salpeter 1997). 
For slowly accreting neutron stars (so that the surface temperature is 
low enough for condensation to occur), the inflowing hydrogen can burn, 
almost as soon as it has condensed into the liquid phase. It is not clear
whether this burning proceeds smoothly or whether there is some kind of 
oscillatory relaxation (e.g., cooling leads to condensation,
leading to hydrogen burning and heat release, followed by evaporation
which stops the burning and leads to further cooling; see Salpeter 1998).

\subsection{Iron Surface Layer}

For neutron stars that have not accreted much gas, one might expect 
the surfaces to consist of iron formed at the neutron star's birth.
As discussed in Sec.~V.E the cohesive energy of Fe is uncertain. 
If the condensed Fe is unbound with respect to the Fe atom ($Q_s=|E_s|-|E_{\rm
atom}|<0$), then the outermost Fe layer of the NS
is characterized by gradual transformation from
nondegenerate gas at low densities, which includes Fe atoms and ions,
to degenerate plasma as the pressure (or column
density) increases. The radiative spectrum will be largely determined by
the property of the nondegenerate layer. However, even a weak cohesion 
of the Fe condensate ($Q_s>0$) can give rise to a phase transition at
sufficiently low temperatures.\footnote{The condensation of Fe was
first discussed by Ruderman and collaborators (see Ruderman 1974;
Flowers \etal~1977), although these earlier calculations greatly overestimated
the cohesive energy $Q_s$ of Fe (see Sec.~V.B and Sec.~V.E).}
The number density of atomic Fe in the 
saturated vapor is of order 
\be
n_A\simeq \left({AMT\over 2\pi}\right)^{3/2}\exp\left(-{Q_s\over T}\right).
\ee
The gas density in the vapor is $\rho_g\go AMn_A$. The critical temperature
for phase transition can be estimated from $\rho_g=\rho_s$. 
Using Eq.~(\ref{rs0}) as an estimate for the condensation density $\rho_s$,
and using Eq.~(\ref{eqqs}) as the upper limit of $Q_s$, we find
\be
T_{\rm crit}\lo 0.1\,Q_s\lo 10^{5.5}\,B_{12}^{2/5}~{\rm K}.
\ee
As in the case of hydrogen (see Sec.~VII.B), we expect the vapor above the
condensed iron surface to have negligible optical depth when $T\lo T_{\rm
crit}/3$. 
 
The iron surface layers of magnetic NSs have also been studied
using Thomas-Fermi type models (e.g., Fushiki, Gudmundsson and Pethick 
1989; Abrahams and Shapiro 1991; R\"ognvaldsson \etal~1993; 
Thorolfsson \etal~1998; see Appendix B). While these models are too crude to 
determine the cohesive energy of the condensed matter, they provide 
a useful approximation to the gross properties of the NS surface layer.
Figure 9 depicts the equation of
state of iron at $B=10^{12}$~G and $B=10^{13}$~G based on Thomas-Fermi
type statistical models. At zero temperature, the pressure is zero 
at a finite density which increases with increasing $B$. 
This feature is qualitatively the same as in the uniform electron gas 
model (Sec.~V.D). Neglecting the exchange-correlation energy and 
the nonuniformity correction, we can write the pressure of a zero-temperature
uniform electron gas as
\be
P=P_e-{3\over 10}\left({4\pi\over 3}\right)^{1/3}\!\!\!\!
(Ze)^2\!\left({\rho\over Am_p}\right)^{4/3},
\label{puniform}\ee
where the first term is given by Eq.~(\ref{eqpe0}) [or by 
Eq.~(\ref{eqpe1}) in the strong field, degenerate limit],
and the second term results from the Couloumb interactions among the 
electrons and ions. Setting $P=0$ gives the condensation density
$\rho_{s,0}$ as in Eq.~(\ref{rs0}). Note that near zero-pressure,
all these models are approximate (e.g., the calculated 
$\rho_{s,0}$ can differ from the true value by a factor of a few), and more
detailed electronic structure calculations are needed to obtain
reliable pressure -- density relation. Figure 9 also shows
the results of the finite-temperature Thomas-Fermi model
(Thorolfsson \etal~1998). Obviously, at finite temperatures, the pressure
does not go to zero until $\rho\rightarrow 0$, i.e., an atmosphere is
present. Note that the finite-temperature Thomas-Fermi
model only gives a qualitative description of the dense 
atmsophere; important features such as atomic states and
ionizations are not captured in this model.

\subsection{Radiative Transfer and Opacities}

As discussed in Sec.~I.A, the surface thermal radiation 
detected from isolated NSs provides valuable information
on the structure and evolution of NSs.
To calculate the radiation spectrum from the NS surface,
one needs to understand the radiative opacities and study
radiative transport in the atmosphere. A thorough review
of these subjects is beyond the scope of this paper. Here we briefly
discuss what has been done and provide a pointer to recent papers.

Nonmagnetic ($B=0$) NS atmosphere models were first constructed by 
Romani (1987). Further works 
(Rajagopal and Romani 1996; Zavlin \etal~1996) used improved opacity and 
equation of state data from the OPAL project (Iglesias and Rogers 1996) for 
pure hydrogen, helium and iron compositions. 
These works showed that the radiation spectra from light-element (hydrogen or 
helium) atmospheres deviate significantly from the blackbody spectrum.
The zero-field models may be relevant for the low-field
($B\sim 10^8-10^9$~G) recycled pulsars (e.g., PSR J0437-4715, from which
thermal X-rays have apparently been detected; see Zavlin and Pavlov 1998), 
but it is possible that these intermediate magnetic fields
can have non-negligible effects, at least for atmospheres
of light elements (e.g., even at $B\sim 10^9$~G, the binding energies
of light elements can differ significantly from the zero-field values, 
and at low temperatures the OPAL equation of state 
underestimates the abundance of the neutral species in the atmosphere).

The basic properties of radiation emerging from a completely ionied
magnetic NS atmosphere (under the assumption of constant temperature gradient
in the raidtaing layer) were considered by Pavlov \& Shibanov (1978).
Magnetic NS atmosphere modeling (which requires determining
the temperature profile in the atmosphere as well as the radiation field 
self-consistently) was first attempted by Miller (1992),
who adopted the polarization-averaged bound-free (photoionization)
opacities (calculated by Miller and Neuhauser 1991) while neglecting other
radiative processes. However, separate transport of polarization modes (which
have very different opacities) dramatically affects the emergent spectral
flux. So far the most comprehensive models of magnetic hydrogen atmospheres
have been constructed by Pavlov and collaborators (e.g., Shibanov \etal~1992;
Pavlov \etal~1994; Zavlin \etal~1995; see Pavlov \etal~1995 for a review). 
These models correctly take account of the transport of different photon modes
through a mostly ionized medium in strong magnetic fields. 
The opacities adopted in the models include free-free
transitions (bremsstrahlung absorption) and electron scattering, while
bound-free (photoionization) opacities are treated in a highly approximate
manner and bound-bound transitions are completely ignored. The models of 
Pavlov \etal are expected to be valid for relatively high temperatures ($T\go
{\rm a~few}\times 10^6$~K) where hydrogen is almost completely ionized. 
As the magnetic field
increases, we expect these models to break down at even higher temperatures as
bound atoms, molecules and condensate become increasingly important. 
The atmosphere models of Pavlov at al. 
have been used to compare with the observed spectra
of several radio pulsars and radio-quiet isolated NSs (e.g., Meyer et
al.~1994; Pavlov \etal~1996; Zavlin, Pavlov and Tr\"umper 1998)
and some useful constraints on the NS properties have been obtained. 
Magnetic iron atmospheres were considered by Rajagopal \etal~(1997),
who adopted an approximate treatment of radiative opacities and transport. 

As discussed in Sec.~VII.A, the strong magnetic field 
increases the abundance of neutral atoms in a hydrogen atmosphere
as compared to the zero-field case (see Fig.~7).
Therefore one could in principle expect some atomic or molecular 
line features in the soft X-ray or UV spectra. For example, 
the Lyman ionization edge is shifted to $160$ eV at $10^{12}$ G 
and $310$ eV at $10^{13}$ G. The free-free and bound-free (for a ground-state
hydrogen atom at rest) cross-sections 
for a photon in the extraordinary mode (with the photon electric field 
perpendicular to the magnetic field) are approximately given by 
(e.g., Gnedin, Pavlov and Tsygan 1974; Ventura \etal~1992)
\footnote{Note that equations (\ref{sigmabfperp}) and
(\ref{sigmabfpara}) are based on the Born approximation,
which breaks down near the photoionization threshold.
A more accurate fitting to $\sigma_{\rm bf}$, valid for any tightly bound
state and photon polarization, is given by eqs.~(39)-(43) of 
Potekhin \& Pavlov (1993).}
\ba
&&\sigma_{{\rm ff}\perp}\simeq 1.7\times 10^3\rho T_5^{-1/2}
\alpha^2a_0^2\omega^{-1}b^{-2},\label{sigmaffperp}\\
&&\sigma_{{\rm bf}\perp}\simeq 4\pi\alpha a_0^2\left({Q_1\over\omega}
\right)^{3/2}\!\!b^{-1},\label{sigmabfperp}
\ea
where $\alpha=1/137$ is the fine structure constant, 
$a_0$ is the Bohr radius, $\rho$ is the density in g$\,$cm$^{-3}$, 
$\omega$ is the photon energy in the atomic units, $Q_1$ is the ionization 
potential, and $b$ is the dimensionless field strength
defined in Eq.~(\ref{eqb0}). Near the absorption edge, the ratio of the 
free-free and bound-free opacities is
\be
{\kappa_{{\rm ff}\perp}\over\kappa_{{\rm bf}\perp}}
\sim 10^{-4}{\rho\over T_5^{1/2}B_{12}f_H}.
\ee
Thus even for relatively small neutral H fraction $f_H$,
the discontinuity of the total opacity at the Lyman edge is 
pronounced and we expect the absorption feature to be prominent. 
Note that since the extraordinary mode has smaller opacity, most of the 
X-ray flux will come out in this mode. For photons in the ordinary 
mode, we have 
\ba
&&\sigma_{{\rm ff}\parallel}\simeq 1.65\times 10^3
\rho T_5^{-1/2}\alpha^2a_0^2\omega^{-3},\label{sigmaffpara}\\
&&\sigma_{{\rm bf}\parallel}\simeq 10^2\pi\alpha
a_0^2(Q_1/\omega)^{5/2}(\ln b)^{-2},
\label{sigmabfpara}\ea
($\sigma_{{\rm ff}\parallel}$ is the same as in zero field),  which give
$\kappa_{{\rm ff}\parallel}/\kappa_{{\rm bf}\parallel} 
\sim 10^{-2}\rho T_5^{-1/2}f_H^{-1}$ for $B_{12}\sim 1$. Thus the absorption 
edge for the ordinary mode is less pronounced.
Clearly, the position and the strength of the absorption 
edge can provide useful diagnostic of the NS surface magnetic field.

A detailed review on electron scattering and free-free absorption
opacities in a magnetized plasma can be found in M\'esz\'aros (1992).
Bound-bound transitions in strong magnetic fields have been 
thoroughly investigated for hydrogen and helium atoms at rest
(see Ruder \etal~1994 for review and tabulations of numerical results).
Bound-free absorptions for hydrogen atoms at rest have also been
extensively studied (e.g., Gnedin \etal~1974; Ventura \etal~1992; Potekhin 
and Pavlov 1993). Since the motion of the atom in strong
magnetic fields modifies the atomic structure significantly (see Sec.~III.E),
the motional effects on opacities need to be included (see
Pavlov and Potekhin 1995 for bound-bound opacities and
Potekhin and Pavlov 1997 for bound-free opacities; see also
Potekhin \etal~1998 for total opacities). 
Some approximate results for radiative transitions in 
heavy atoms were presented by Miller and Neuhauser (1991). 

In the regime where the outermost layer of the NS is in the formed
of condensed matter, radiation directly emerges from the hot (but degenerate)
condensed phase. The usual radiative transfer does not apply to this
situation. Because of the high condensation density, the electron plasma
frequency is larger than the frequency of a typical thermal photon. Thus
the photons cannot be easily excited thermally inside 
the condensate. This implies that the NS surface has a high reflectivity. 
The spectral emissivity is determined by the bulk dielectric properties
of the condensed phase (see Itoh 1975; Brinkmann 1980).

\subsection{Magnetized Neutron Star Crust}

As discussed in Sec.~VI, the effect of 
Landau quantization is important only for 
$\rho\lo \rho_B$ [see Eq.~(\ref{eqrhob})]. Deeper in the NS
envelope (and the interior), we expect the magnetic field effect on the bulk
equation of state to become negligible as more any more Landau levels are
filled. In general, we can use the condition $\rho_B\go \rho$, or $B_{12}\go
27\,(Y_e\rho_6)^{2/3}$, to estimate the critical value of $B$ above which
Landau quantization will affect physics at density $\rho$. For example, 
at $B\go 10^{14}$~G, the neutronization transition from $^{56}$Fe 
to $^{62}$Ni (at $\rho=8.1\times 10^{6}$~g~cm$^{-3}$ for $B=0$; Baym, Pethick
and Pines 1971) in the crust can be significantly affected by the magnetic 
field (Lai and Shapiro 1991).

The ions in the NS envelope form a one-component plasma
and are characterized by the Coulomb coupling parameter
\be
\Gamma={(Ze)^2\over r_i kT}=22.75\,{Z^2\over T_6}\left({\rho_6\over A}
\right)^{1/3},
\ee
where $r_i=(3/4\pi n_i)^{1/3}$ is the Wigner-Seitz cell radius, $n_i=\rho/m_i
=\rho/(Am_p)$ is the ion number density, $\rho_6=\rho/(10^6\,{\rm g~cm}^{-3})$
and $T_6=T/(10^6\,{\rm K})$. For $\Gamma\ll 1$, the ions form
a classical Boltzmann gas whose thermodynamic property is unaffected
by the magnetic field. For $\Gamma\go 1$, the ions constitute
a strongly coupled Coulomb liquid. The liquid freezes into a Coulomb crystal
at $\Gamma=\Gamma_m\simeq 175$, corresponding to the classical melting
temperature $T_m$ (e.g., Slattery \etal~1980; Nagara \etal~1987; Potekhin \&
Chabrier 2000).
The quantum effects of ion motions (zero-point
vibrations) tends to increase $\Gamma_m$ (e.g., Chabrier,
Ashcroft and DeWitt 1992; Chabrier 1993) 
or even suppress freezing (e.g., Ceperley and Alder
1980; Jones and Ceperley 1996). At zero-field, the ion zero-point vibrations
have characteristic frequency of order the ion plasma frequency $\Omega_p$,
with 
\be
\hbar\Omega_p=\hbar\left({4\pi Z^2e^2n_i\over m_i}\right)^{1/2}=675\,
\left({Z\over A}\right)\,\rho_6^{1/2}~{\rm eV}.
\ee
For $T\ll T_{\rm Debye}\sim \hbar\Omega_p/k$, the ion vibrations are 
quantized. The effects of magnetic field on strongly coupled Coulomb liquids
and crystals have not been systematically studied (but see Usov \etal~1980).
The cyclotron frequency of
the ion is given by
\be
\hbar\omega_{ci}=\hbar{ZeB\over Am_pc}=6.3\,\left({Z\over A}\right)\,B_{12}~
{\rm eV}.
\ee
The ion vibration frequency in a magnetic field may be estimated as
$(\Omega_p^2+\omega_{ci}^2)^{1/2}$. Using Lindeman's rule,
we obtain a modified melting criterion:
\be
\Gamma\,\left(1+{\omega_{ci}^2\over\Omega_p^2}\right)\simeq 175.
\ee
For $\omega_{ci}\ll \Omega_p$, or $B_{12}\ll 100\,\rho_6^{1/2}$, the magnetic
field does not affect the melting criterion and other properties of 
ion vibrations.

Even in the regime where the magnetic quantization effects 
are small ($\rho\gg\rho_B$),
the magnetic field can still strongly affect 
the transport properties (e.g., electric conductivity and heat conductivity)
of the NS crust. This occurs when the effective
gyro-freqyency of the electron, $\omega_{ce}^\ast=eB/(m_e^\ast c)$, 
where $m_e^\ast=\sqrt{m_e^2+(p_F/c)^2}$, is much
larger than the electron collision frequency, i.e.,
\be
\omega_{ce}^\ast\tau_0\simeq
1.76\times 10^3\,{m_e\over m_e^\ast}\,B_{12}\,\left({\tau_0\over
10^{-16}\,{\rm s}}\right),
\ee
where $\tau_0$ is the effective electron relaxation time 
($10^{-16}$~s is a typical value in the outer crust). For example,
when $\omega_{ce}^\ast\tau_0\gg 1$, the electron heat conductivity
perpendicular to the magnetic field is suppressed by a factor 
$(\omega_{ce}^\ast\tau_0)^{-2}$. A detailed review on the transport
properties of the magnetized NS crust is given in 
Hernquist (1984) and Yakovlev and Kaminker (1994) where many earlier
references can be found (see Potekhin 1999 for a recent calculation). 
The thermal structure of a magnetized neutron star crust 
has been studied by, e.g., Hernquist (1995), Van Riper (1988), Schaaf (1990),
Heyl and Hernquist (1998).

\section{Concluding Remarks}

The properties of matter in strong magnetic fields has always been 
an interesting subject for physicists. While early studies
(e.g., Elliot and Loudon 1960; Hasegawa and Howard 1961)
were mainly motivated by the fact that the strong magnetic field
conditions can be mimicked in some semiconductors where a small
effective electron mass and a large dielectric constant reduce the electric
force relative to the magnetic force, recent work on this
subject has been motivated by the huge magnetic field $\sim 10^{12}$~G
known to exist in many neutron stars and the tentative evidence
for fields as strong as $10^{15}$~G (see Sec.~I.A).
The study of matter in strong magnetic fields 
is obviously an important component of neutron star astrophysics research.
In particular, interpretation of the ever improving spectral 
data of neutron stars requires a detailed theoretical understanding 
of the physical properties of highly-magnetized atoms, molecules, and
condensed matter.

In this review, we have focused on the electronic structure and 
the bulk properties of matter in strong magnetic fields. We have only 
briefly discussed the issues of radiative opacities and conductivities
in a magnetized medium (see Sec.~VII.D-E). There are other related 
problems that are not covered in this paper. For example, 
neutrino emissions in the neutron star crust and interior, 
which determine the cooling rate of the star, can be affected by strong
magnetic fields (e.g., Yakovlev and Kaminker 1994; Baiko and Yakovlev 1999;
van Dalen \etal~2000 and references therein). In proto-neutron stars,
sufficiently strong magnetic fields ($B\go 10^{15}$~G) can induce asymmetric
neutrino emission and impart a kick velocity to the star (e.g.,
Dorofeev \etal~1985; Vilenkin 1995; Horowitz and Li 1998; 
Lai and Qian 1998a,b; Arras and Lai 1999a,b and references therein). 
We have not discussed the magnetic field effects on the 
nuclear matter equation of
state (e.g., Chakrabarty \etal~1997; Yuan and Zhang 1999; 
Broderick \etal~2000; Suh \& Mathews 2000), 
which are relevant for $B\go 10^{17}-10^{18}$~G. Many other
aspects of physics in strong magnetic fields are reviewed in Meszaros (1992).

The study of matter in strong magnetic fields spans a number of different 
subareas of physics, including astrophysics, atomic and molecular physics,
condensed matter physics and plasma physics. 
As should be apparent from the preceding sections, although there has
been steady progress over the years, many open problems remain to be studied 
in the future. Some of the problems should be easily solvable by specialists
in their respective subareas. 
Since there is no recent review that covers this broad subject area,
our emphasis has been on producing a self-contained account on 
the basic physical pictures, while relegating the details to the original
literature (although we have also discussed aspects of calculational 
techniques). We hope that this review will
make the original literature on matter in strong magnetic fields 
more easily accessible and stimulate more physicists to work on this problem.

\section*{ACKNOWLEDGMENTS}

I am grateful to Edwin Salpeter for providing many insights
on the subject of this paper 
and for his advice on writing this article. 
I thank George Pavlov and Alexander Potekhin 
for detailed comments and suggestions.
I also thank Wynn Ho, Edwin Salpeter and Ira Wasserman for commenting on an
early draft of this paper, 
as well as Elliot Lieb, Peter Schmelcher and Jokob Yngvason for
useful comments/communications. 
This work has been
supported in part by NASA grants NAG 5-8484 and NAG 5-8356, by NSF grant AST
9986740, and by a research fellowship from the Alfred P. Sloan foundation.   

\appendix

\section{Jellium Model of Electron Gas in Strong Magnetic Fields}

Consider an electron gas in a uniform background of positive
charges (the ``Jellium model''). The interelectronic spacing $r_s$ is related
to the electron number density $n_e$ by $n_e^{-1}=4\pi r_s^3/3$. 
At zero magnetic field, the energy per electron can be written as 
(e.g., Ashcroft and Mermin 1976)
\be
E={3\over 5}{\hbar^2k_F^2\over 2m_e}-{3\over 4}{e^2k_F\over\pi}
+E_{\rm corr}
={2.21\over r_s^2}-{0.916\over r_s}+0.0622\ln(r_s)-0.094~({\rm Ryd}).
\ee
The first term on the right-hand side 
is the kinetic energy, the second term the (Hartree-Fock)
exchange energy, and the remaining terms are the correlation energy 
(the expression for the correlation energy given here 
applies to the $r_s\ll 1$ limit; see Ceperley and Alder 1980 for the full 
result).

Now consider the magnetic case. When the density is low (or the
magnetic field high) so that only the ground Landau level is occupied, the
Fermi wavenumber $k_F$ can be calculated from
\be
n_e={eB\over hc}{1\over 2\pi}\int_{-k_F}^{k_F}dk_z={2k_F\over
(2\pi\hat\rho)^2}\Longrightarrow k_F=2\pi^2\hat\rho^2n_e.
\ee
It is convenient to introduce the new parameters, the inverse Fermi wavenumber
$r_F$ and the filling factor for the lowest Landau level $t$,
\be 
r_F={1\over\pi k_F}={2\over 3\pi^2}{r_s^3\over\hat\rho^2},\qquad
t={\eps_F\over\hbar\omega_{ce}}=\left({n_e\over n_{B}}\right)^2=
{9\pi^2\over 8}\left({\hat\rho\over r_s}
\right)^6,
\ee
where $\eps_F=(\hbar k_F)^2/(2m_e)$ is the Fermi energy. For the electrons
to occupy only the ground Landau level we require $t\le 1$, or $n_e
\le n_{B}$ [see Eq.~(\ref{nlandau})]. The kinetic energy per electron is
\be
\eps_k={1\over n_e}{1\over (2\pi\hat\rho)^2}\int_{-k_F}^{k_F}{\hbar^2k_z^2\over
2m_e}\,dk_z
={\hbar^2\over 6\pi^2m_er_F^2}.
\label{eqek}\ee
The (Hartree-Fock) exchange energy can be written as (Danz and Glasser 1971)
\be
\eps_{ex}=-{F(t)e^2\over 2\pi^2r_F},
\ee
where the dimensionless function $F(t)$ is 
\be
F(t)=4\int_0^\infty\!\!\!dx\left[\tan^{-1}\left({1\over
x}\right)-{x\over 2}\ln\left(1+{1\over x^2}\right)\right] e^{-4tx^2}.
\ee
The function $F(t)$ can be expressed in terms of generalized hypergeometric
functions; for small $t$, it can be expanded as (Fushiki \etal~1989)
\be
F(t)=3-\gamma-\ln(4t)+{2t\over 3}\left({13\over 6}-\gamma-\ln 4t\right)
+{8t^2\over 15}\left({67\over 30}-\gamma-\ln 4t\right)+{\cal O}(t^3\ln t),
\ee
where $\gamma=0.5772\cdots$ is Euler's constant.
The expressions of exchange energy for higher Landau levels are
given in Fushiki \etal~(1992). 

Correlation energy in strong magnetic fields has been calcualated by 
Steinberg and Ortner (1998) in the limit of $r_F\ll 1$ and $t\ll 1$
in the random phase approximation (see also Skudlarski and Vignale 1993).
The leading order terms have the
form 
\be
\eps_{\rm corr}=D(t)\ln r_F+C(t),
\ee
with $D(t)=(16\pi^2t)^{-1}$ (Horing \etal~1972); asymptotic ($t\ll 1$)
expression for $C(t)$ is given in Steinberg and Ortner (1998). 
There are also some studies (e.g., Kleppmann and Elliot 1975; 
Usov \etal~1980; MacDonald and Bryant 1987) for the very low density regime
($r_F\gg 1$, but $t<1$), where the electron gas is expected to form a Wigner
crystal; a strong magnetic field tends to increase the density at which
crystalization occurs. The low-density and high-density limits may be used to
establish an interpolation formula for the correlation energy.

\section{Thomas-Fermi Models in Strong Magnetic Fields}

Thomas-Fermi (TF) and related theories (e.g., Thomas-Fermi-Dirac theory,
which includes the exchange energy of electrons)
have been thoroughly studied as models of atoms and bulk matter 
in the case of zero magnetic field (e.g., Lieb 1981; 
Spruch 1991). Since the early studies of Kadomtsev (1970) and Mueller, Rau and
Spruch (1971), there has been a long succession of papers on TF type
models in strong magnetic fields.
An excellent review on the subject is given in Fushiki \etal~(1992); 
rigorous theorems concerning the validity of TF theory in magnetic fields 
as an approximation of quantum mechanics are discussed in 
Yngvason (1991), Lieb \etal~(1992,1994a,b)
and references therein. Here we briefly summarize the basics of the theory.

The TF type theory is the simplest case of a density functional theory.
For a spherical atom (or a Wigner-Seitz cell of bulk matter), we write
the energy as a functional of the local electron density $n(r)$:
\be
\cE[n(r)]=\int\!\!w_e[n(r)]\,d^3r-Ze^2\int\!{n(r)\over r}\,d^3r
+{e^2\over 2}\int\!{n(r)n(r')\over |\br-\br'|}\,d^3r d^3r',
\ee
where $w_e[n(r)]$ is the energy density of electrons at density $n(r)$, which
includes kinetic, exchange, and correlation energy (see Appendix A).  
Minimizing $\cE$ with respect to $n(r)$ yields
the TF equation:
\be
{dw_e\over dn}-e\Phi(r)=\mu_{\rm TF},
\label{eqthomas}\ee
where $\Phi(r)$ is the total electrostatic potential, and
$\mu_{\rm TF}$ is a constant to be determined by the constraint
$\int\!n(r)\,d^3r=Z$ (for neutral system). Equation (\ref{eqthomas})
can be solved together with the Poisson equation
\be
\nabla^2\Phi=-4\pi Ze\delta(\br)+4\pi en(r),
\label{eqpoisson}\ee
to obtain $n(r)$ and $\Phi(r)$, and the energy $\cE$ can then be evaluated.
The pressure of the bulk matter is then given by
$P=-d\cE/(4\pi r_i^2 dr_i)$ and is equal to the pressure of
a uniform electron gas at density $n(r_i)$ (where $r_i$ is the Wigner-Seitz 
radius). 

Consider the TF model, in which we include in $w_e[n(r)]$ only the kinetic
energy of the electrons. We then have $dw_e/dn=\mu_e(r)=e\Phi(r)+\mu_{\rm TF}$.
Using Eq.~(\ref{eqne2}) (for nonrelativistic electrons) we can express
the local electron density $n(r)$ in terms of $\Phi(r)$. At zero temperature,
we have
\be
n(r)={1\over \sqrt{2}\pi^2\hat\rho^3}\sum_{n_L}^{n_{\rm max}}
g_{n_L}\!\left[{\mu_{\rm TF}+e\Phi(r)\over \hbar\omega_{ce}}-n_L\right]^{1/2}.
\ee
This can be substituted into Eq.~(\ref{eqpoisson}) to solve for $\Phi(r)$.

Many TF type studies (e.g., Kadomtsev 1970; Mueller, Rau and Spruch 1971; 
Banerjee \etal~1974; Constantinescu and Reh\'ak 1976; Skjervold and
\"Ostgaard 1984; Fushiki \etal~1989; Abrahams and Shapiro 1991)
adopted the adiabatic approximation (only the ground Landau level is 
occupied, i.e., $n_{\rm max}=0$). Abrahams and Shapiro (1991) also considered
the Weizs\"acker gradient correction where a term of the form
$(\hbar^2/m_e)(\nabla n)^2/n$ is included in $w_e[n]$, although
the actual form of the Weizs\"acker term in magnetic fields is uncertain
(see Fushiki \etal~1992 for a critical review).
Tomishima \& Yonei (1978), Tomishima \etal~(1982), Gadiyak \etal~(1981), 
and R\"ognvaldsson \etal~(1993)
studied TF models allowing for $n_{\rm max}>0$. 
Constantinescu and Moruzzi (1978), Abrahams and Shaprio (1991)
(who restricted to $n_{\rm max}=0$) and Thorolfsson \etal~(1998) (allowing
for $n_{\rm max}>0$) studied the finite temperature TF models for condensed 
matter. Examples of TF type equations of state for iron are shown in Fig.~9.

As mentioned, the TF model is too crude to determine the relative binding
between atoms, molecules and condensed matter. But the model becomes
increasingly accurate in certain asymptotic limits (see
Fushiki \etal~1992; Lieb \etal~1992,1994a,b). In general, the validity of
the TF theory requires the local Fermi wavelength $\lambda_F$ to be much 
less than the size of the atom (or Wigner-Seitz cell) $r_i$. 
For zero or weak magnetic fields ($n_{\rm max}\gg 1$), $\lambda_F\sim 
n_e^{-1/3}\sim r_i/Z^{1/3}$, the validity of TF theory requires
$Z^{1/3}\gg 1$. The size of the atom in weak fields is estimated from
$\hbar^2/(m_e\lambda_F^2)\sim Ze^2/r_i$, which gives $r_i\sim Z^{-1/3}a_0$.

For atoms in strong magnetic fields ($n_{\rm max}=0$), we have $\lambda_F
\sim r_i^2/(Z\hat\rho^2)$, and the size is $r_i\sim Z^{1/5}b^{-2/5}a_0$.
The condition $\lambda\ll r_i$ becomes $b\ll Z^3$. The TF theory of the atom is
exact when $Z,b\rightarrow\infty$, as long as $Z^3/b\rightarrow\infty$.
To be in the strong field regime requires the mean electron
spacing in a nonmagnetic atom, $a_0/Z^{2/3}$, to be greater than $\hat\rho$,
i.e., $b>Z^{4/3}$.
For bulk matter in strong magnetic fields ($n_{\rm max}=0$), 
the TF model is valid when $\lambda_F\sim r_i^3/(Z\hat\rho^2)\ll r_i$,
i.e., $b\ll Zr_i^{-2}$ (at zero-pressure, $r_i\sim Z^{1/5}b^{-2/5}$, this
condition reduces to $b\ll Z^3$). The strong field condition is
$Z^{-1/3}r_i>\hat\rho$, i.e., $b>Z^{2/3}r_i^{-2}$ (this reduces to
$b>Z^{4/3}$ at zero-pressure density).



\newpage

\begin{figure}[t]
\includegraphics{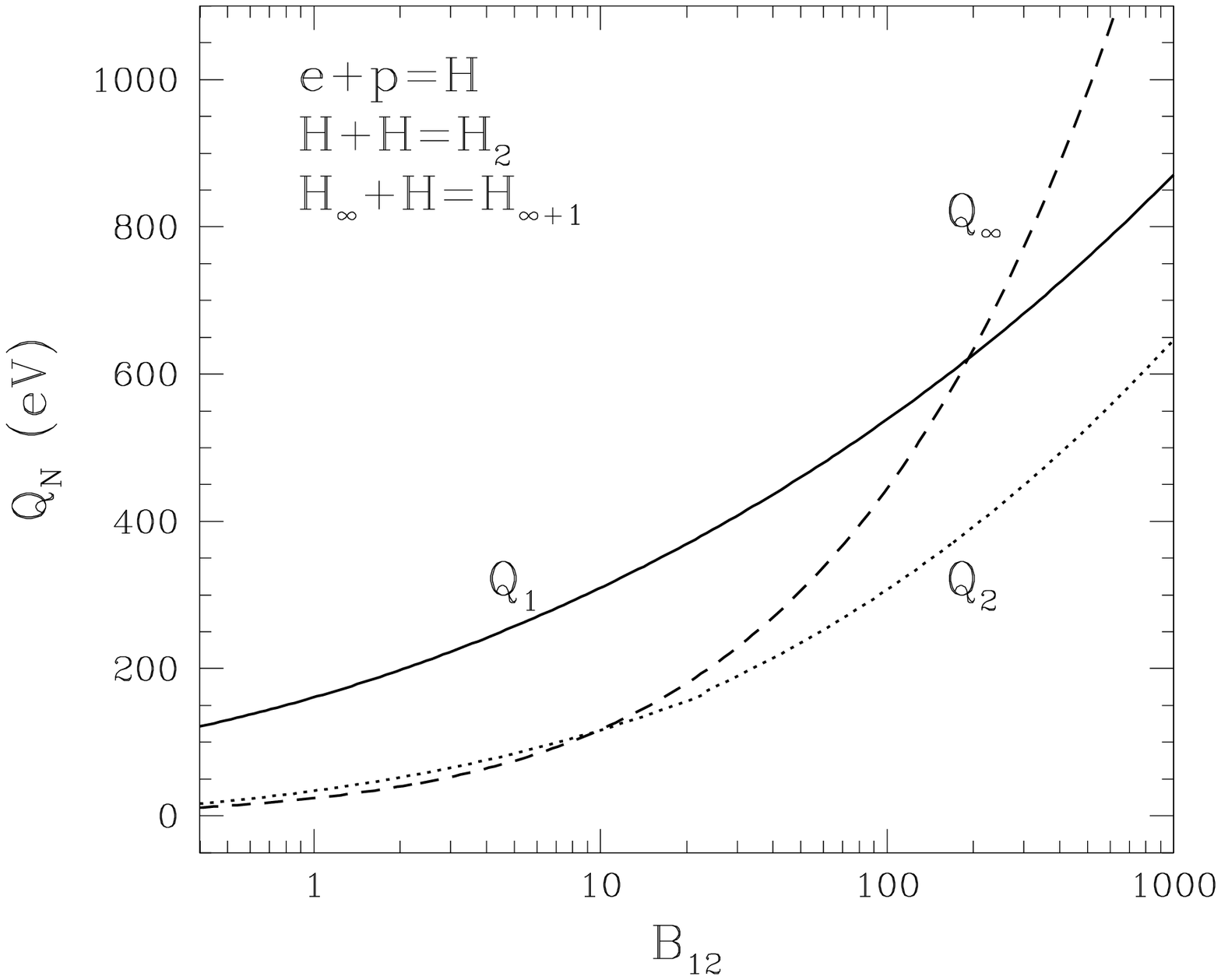}
\vspace*{60mm}
\caption{Energy releases from several atomic and molecular processes as a
function of the magnetic field strength. The solid line shows the ionization
energy $Q_1$ of H atom, the dotted line shows the dissociation energy $Q_2$ of
H$_2$, and the dashed line shows the cohesive energy $Q_\infty$ of linear 
chain H$_\infty$. The zero-point energy corrections have been included in
$Q_2$ and $Q_\infty$.}
\end{figure}

\begin{figure}[t]
\includegraphics{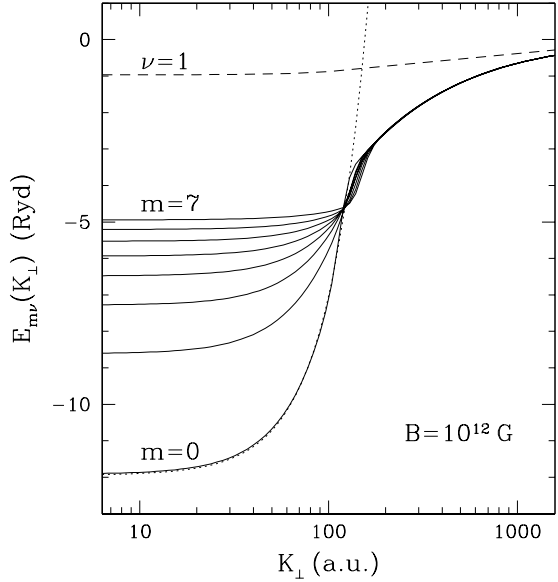}
\vspace*{60mm}
\caption{Energy spectrum of a hydrogen atom moving across a magnetic
field with $B=10^{12}$~G. The solid lines are for the tightly bound
states ($\nu=0$) with $m=0,1,2,\cdots,7$, and the dashed line is
for weakly bound state with $\nu=1,\,m=0$. The dotted line corresponds
to $E_{00}(K_\perp)=E_{00}(0)+K_\perp^2/(2M_\perp)$. The total energy of
the atom is ${\cal E}_{m\nu}=K_z^2/(2M)+m\hbar\omega_{cp}+E_{m\nu}(K_\perp)$.}
\end{figure}

\clearpage

\begin{figure}[t]
\includegraphics{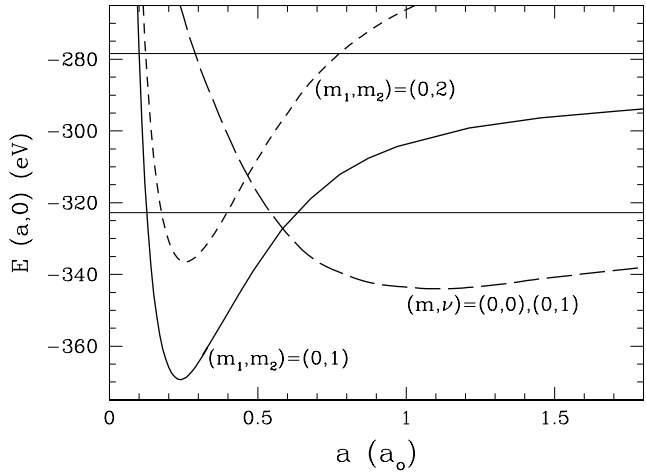}
\vspace*{50mm}
\caption{The electronic energy curves $E(a,0)$ of H$_2$ molecule 
at $B=10^{12}$~G when the molecular axis is aligned with 
the magnetic field axis ($a$ is the proton separation). 
For the solid curve the electrons occupy the 
$m_1=0$ and $m_2=1$ orbitals (both with $\nu=0$), for the short-dashed
curve $(m_1.m_2)=(0,2)$. The long-dashed curve corresponds to the weakly 
bound state with $(m,\nu)=(0,0),(0,1)$. The solid horizontal lines
correspond to $E=-323$~eV (the total energy of two isolated atoms in 
the ground state) and $E=-278$~eV [the total energy of two isolated
atoms, one in the ground state ($-161$~eV), another in the first excited
state ($-117$~eV)].
}
\end{figure}

\begin{figure}[t]
\includegraphics{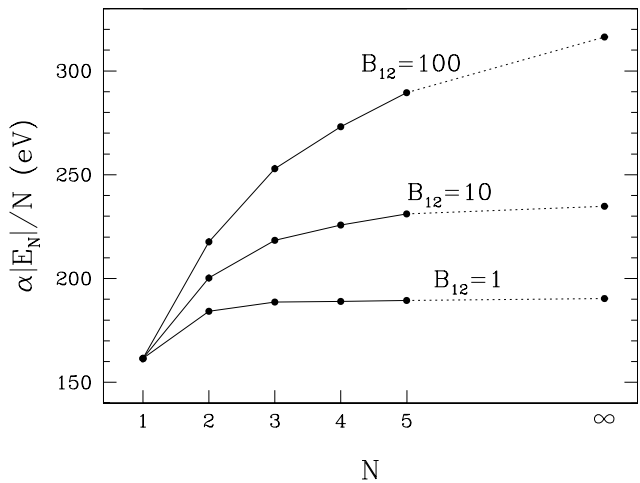}
\vspace*{60mm}
\caption{Binding energy per atom, $|E_N|/N$, for H$_N$ molecules 
in strong magnetic fields as a function of $N$. To facilitate 
plotting, the values of $|E_1|$ at different $B_{12}$ are normalized to 
its value ($161.5$~eV) at $B_{12}=1$; This means $\alpha=1$ for 
$B_{12}=1$, $\alpha=161.5/309.6$ for $B_{12}=10$ and
$\alpha=161.5/541$ for $B_{12}=100$.
}
\end{figure}

\clearpage

\begin{figure}[t]
\includegraphics{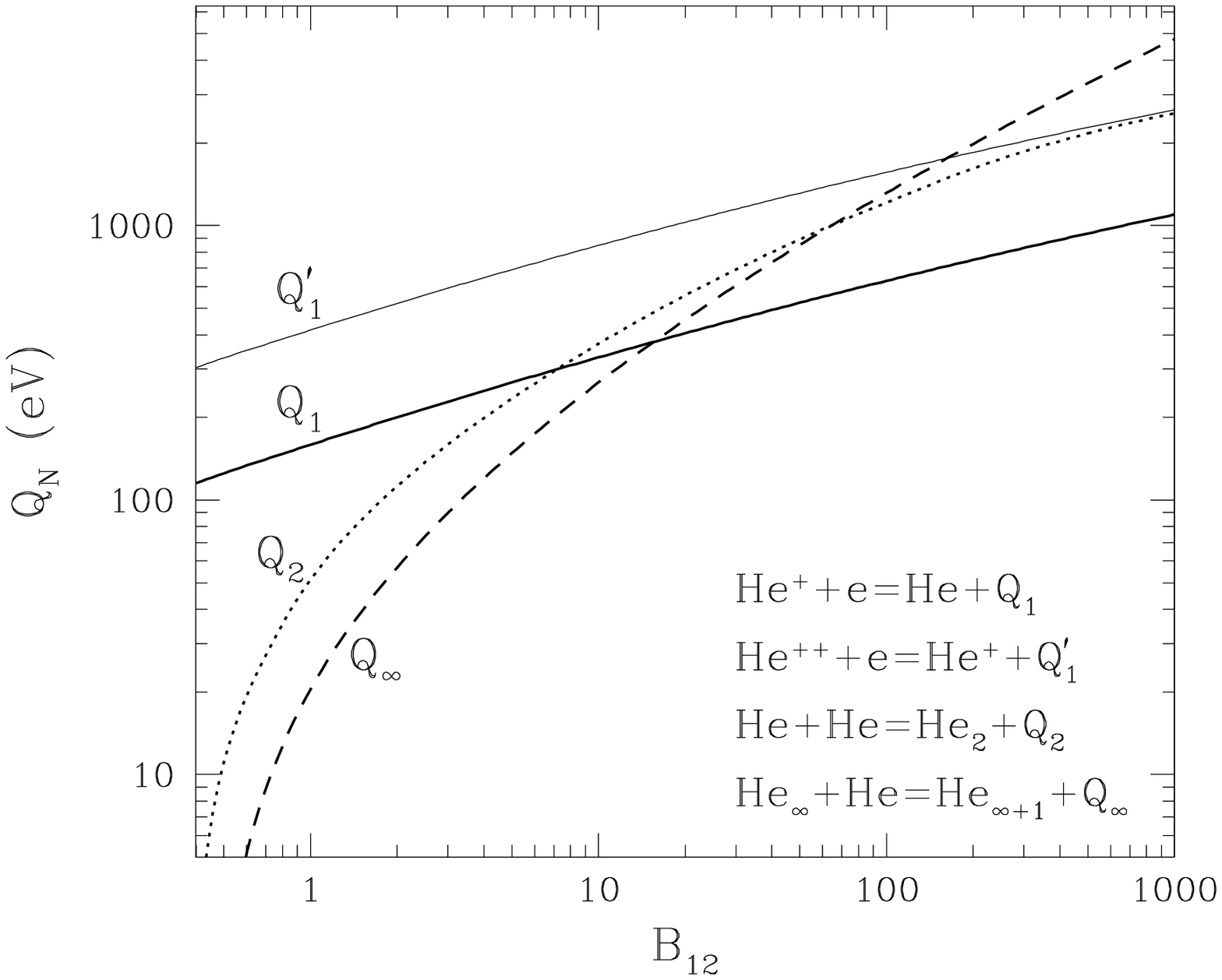}
\vspace*{60mm}
\caption{Energy releases from several atomic and molecular processes as a
function of the magnetic field strength. The solid line shows the ionization
energy $Q_1$ of He atom, the dotted line shows the dissociation energy $Q_2$ of
He$_2$, and the dashed line shows the cohesive energy $Q_\infty$ of linear 
chain He$_\infty$. The zero-point energy corrections are not included.
}
\end{figure}

\begin{figure}[t]
\includegraphics{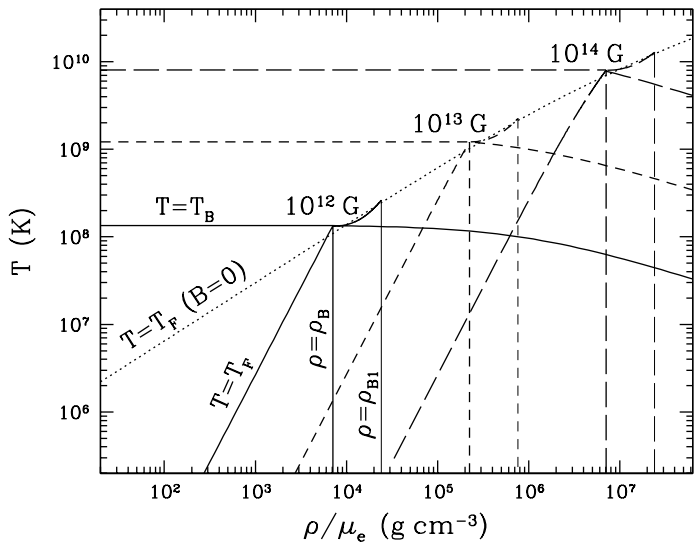}
\vspace*{60mm}
\caption{Temperature-density diagram illustrating the different regimes 
of magnetic field effects on the thermodynamic properties of 
a free electron gas. The solid lines are for $B=10^{12}$~G,
short-dashed lines for $B=10^{13}$~G, and long-dashed lines for 
$B=10^{14}$~G. For each value of $B$, the vertical lines
correspond to $\rho=\rho_B$ (the density below which only the ground Landau
level is occupied by the degenerate electrons) and $\rho=\rho_{B1}$ (the
density below which only the $n_L=0,1$ levels are occupied); the Fermi
temperature is shown for $\rho\le\rho_B$ and for $\rho_B<\rho\le\rho_{B1}$;
the line marked by ``$T=T_B$'' [see Eq.~(\ref{eqtb})] corresponds to
the temperature above which the Landau level effects are smeared out.
The dotted line gives the Fermi
temperature at $B=0$. The magnetic field is strongly quantizing when
$\rho\lo \rho_B$ and $T\lo T_B$, weakly quantizing when $\rho\go \rho_B$ and
$T\lo T_B$, and non-quantizing when $T\gg T_B$. See Sec.~VI for detail.   
}
\end{figure}

\clearpage

\begin{figure}[t]
\includegraphics{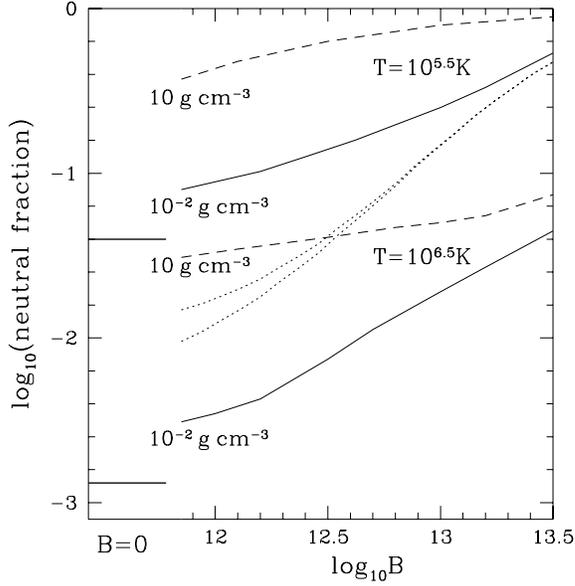}
\vspace*{60mm}
\caption{Atomic hydrogen fraction $n_H/n_b$ as a function of 
magnetic field strength. The solid lines (for $T=10^{5.5}$~K 
and $T=10^{6.5}$~K) are for density $\rho=0.01$~g~cm$^{-3}$, and the dashed
lines for $\rho=10$~g~cm$^{-3}$. These results are based on the calculations
of Potekhin \etal~(1999). The two dotted lines (for $T=10^{5.5}$~K,
$\rho=0.01$~g~cm$^{-3}$) correspond to the hydrogen fraction $n_H'/n_b$
calculated ignoring the ``decentered states'' (the lower line includes 
only the $m=0$ state, while the upper line includes all $m$-states). 
The two horizontal lines on the left correspond to zero-field results
(for $\rho=0.01$~g~cm$^{-3}$ and $T=10^{5.5},10^{6.5}$~K;
for $\rho=10$~g~cm$^{-3}$, all atoms are pressure-ionized at zero field).
}
\end{figure}

\begin{figure}[t]
\includegraphics{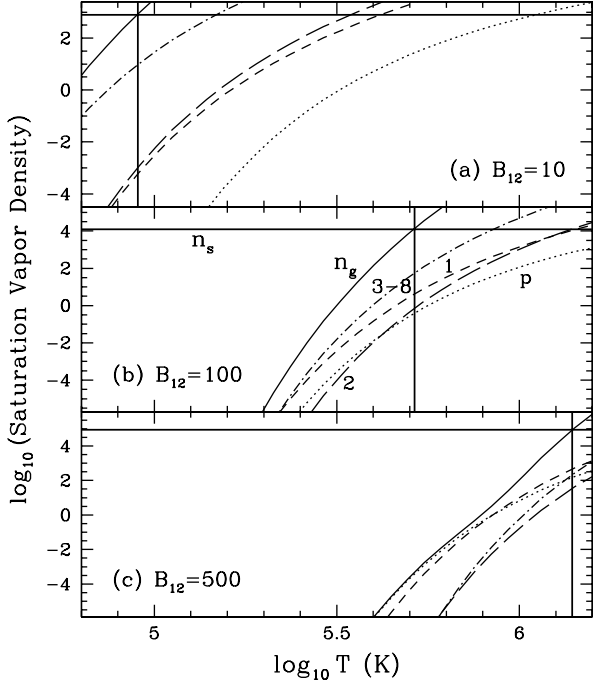}
\vspace*{78mm}
\caption{
The saturation vapor densities of various species 
(in the atomic units, $a_0^{-3}$) of condensed metallic hydrogen 
as a function of temperature for different magnetic 
field strengths: (a) $B_{12}=10$; (b) $B_{12}=100$; (c) $B_{12}=500$.
The dotted curves give $n_p$, the short-dashed curves give $n(\rH)$,
the long-dashed curves give $n(\rH_2)$, the dot-dashed curves
give $[3n(\rH_3)+4n(\rH_4)+\cdots+8n(\rH_{8})]$, and the
solid curves give the total baryon number density in the vapor
$n_g=n_p+n(\rH)+2n(\rH_2)+\cdots$. 
The horizontal solid lines denote the condensation density $n_s\simeq
50\,B_{12}^{6/5}$ (a.u.), while the vertical solid lines correspond to 
the critical condensation temperature at which $n_g=n_s$ (From 
Lai and Salpeter 1997).
}
\end{figure}

\clearpage

\begin{figure}[t]
\includegraphics{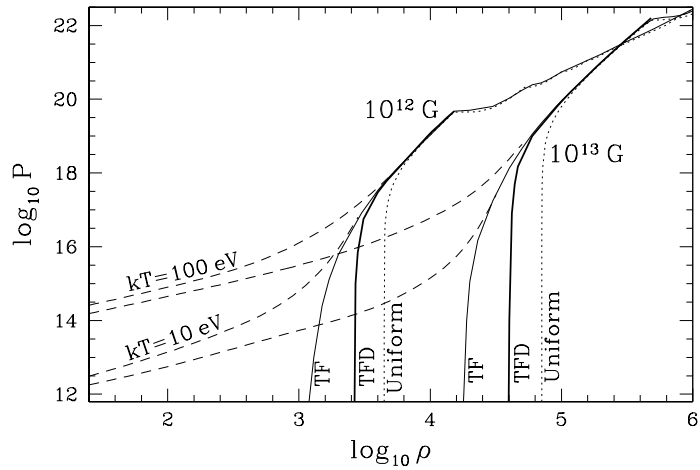}
\vspace*{60mm}
\caption{Equation of state of Fe in strong magnetic fields 
($B=10^{12},10^{13}$~G) based on Thomas-Fermi type models.
The light solid lines are results of Thomas-Fermi (TF) model allowing for many
Landau levels (R\"ognvaldsson \etal~1993), the thick solid lines are results
of Thomas-Fermi-Dirac model allowing only for the ground Landau level (Fushiki
\etal~1989), and the dotted lines are results of the uniform gas model as
descibed by Eq.~(\ref{puniform}), 
all for zero temperature. The dashed lines are results
of Thomas-Fermi models with $kT=10,~100$~eV (Thorolfsson \etal~1998). 
}
\end{figure}

\clearpage
\end{document}